\documentclass[epjST]{svjour}

\usepackage{amsmath}
\usepackage{amssymb,amsfonts,latexsym}
\usepackage{bm}
\usepackage[mathcal]{euscript}
\usepackage{graphicx}
\usepackage{epsfig}
\usepackage{color}

\newcommand{\beq}{\begin{equation}}
\newcommand{\eeq}{\end{equation}}

\begin{document}

\title{Boltzmann-Ginzburg-Landau approach for continuous descriptions of generic Vicsek-like models}

\author{Anton Peshkov\inst{1} \and Eric Bertin\inst{2,3} \and Francesco Ginelli\inst{4} \and Hugues Chat\'{e}\inst{5,6}}

\institute{Physique et M\'ecanique des Milieux H\'et\'erog\`enes, CNRS UMR 7636, Ecole Sup\'erieure de Physique et de Chimie Industrielles, 10 rue Vauquelin, 75005 Paris, France
\and Laboratoire Interdisciplinaire de Physique, Universit\'e Joseph Fourier Grenoble, CNRS UMR 5588, BP 87, 38402 Saint-Martin d'H\`eres, France
\and Universit\'e de Lyon, Laboratoire de Physique, ENS Lyon, CNRS, 46 all\'ee d'Italie, 69007 Lyon, France
\and SUPA, Institute for Complex Systems and Mathematical Biology,
King's College, University of Aberdeen, Aberdeen AB24 3UE, United Kingdom
\and Service de Physique de l'Etat Condens\'e, CNRS URA 2464, CEA-Saclay, 91191 Gif-sur-Yvette, France
\and LPTMC, CNRS UMR 7600, Universit\'e Pierre et Marie Curie, 75252 Paris, France}

\authorrunning{A. Peshkov et.al.}

\titlerunning{Continuous descriptions of generic Vicsek-like models}

\date{\today}

\abstract{We describe a generic theoretical framework, denoted as the Boltzmann-Ginzburg-Landau approach, to derive continuous equations for the polar and/or nematic order parameters describing the large scale behavior of assemblies of point-like active particles interacting through polar or nematic alignment rules. Our study encompasses three main classes of dry active systems, namely polar particles with 'ferromagnetic' alignment (like the original Vicsek model), nematic particles with nematic alignment ("active nematics"), and polar particles with nematic alignment ("self-propelled rods"). The Boltzmann-Ginzburg-Landau approach combines a low-density description in the form of a Boltzmann equation, with a Ginzburg-Landau-type expansion close to the instability threshold of the disordered state. We provide the generic form of the continuous equations obtained for each class, and comment on the relationships and differences with other approaches.}

%

\maketitle

\section{Introduction}

Active matter is currently a very... active field. Under this vocable one usually has in mind collections of ``active units"
able to extract and dissipate energy from their surrounding to produce systematic and often persistent
motion \cite{SR-REVIEW}. 
The far-from-equilibrium collective dynamics of systems as
diverse as vertebrate groups \cite{Couzin}, insects swarms \cite{Buhl}, colonies of bacteria
\cite{Swinney}, motility assays \cite{Schaller,Sumino2012}, as well
as driven granular matter \cite{Drods,DYCOACT} are thus nowadays routinely modeled by 
collections of locally interacting active particles.

Many such models have now been proposed and studied with various degrees of sophistication and intended realism, 
but the Vicsek model stands out for both its simplicity and its historical role in marking the irruption of physicists in collective motion 
studies. In 1995, Vicsek {\it et al.} considered constant-speed point particles aligning with neighbors in the presence of noise. 
This driven-overdamped, first-order dynamical rule, in spite of its minimality, bears some relevance in all situations where the fluid surrounding the particles can be safely
neglected, e.g. when the particles are moving on a substrate acting as a momentum sink. Many of the examples listed above fall into
this category of ``dry active matter". 

Even within the seemingly restrictive setting of competition between (effective) alignment and noise, a wealth of 
important results has been gathered over the recent years, the most generic ones being the presence of 
long-range correlations and anomalously-strong ``giant" number fluctuations in homogeneous, orientationally ordered, 
active phases, and the spontaneous emergence of large-scale high-density/high order structures in the region bordering the 
onset of order.
Furthermore, a picture of three possible classes of models has been suggested, defined by the symmetries of particles 
and alignment rules. The ``polar class" is that of the original Vicsek model: it deals with polar particles aligning ferromagnetically. 
In ``active nematics", particle carrying a uniaxial director move back and forth along it and align nematically. 
Finally, polar particles aligning nematically can be thought of as self-propelled ``rods". 

Continuous theories have been proposed for these three classes, aiming to
capture the long-wavelength behavior of the
 slow modes of the dynamics. This program was first carried over
by Toner and Tu \cite{TT,TTR-review} who formulated a phenomenological hydrodynamic
theory for the polar class based on conservation laws and symmetry
considerations. Extension to active nematics was
performed in Ref. \cite{GNF1} along the same lines.
While successful at describing fluctuations of homogeneous
ordered phases, this
phenomenological approach is of course unable to build accurate connections to particle-based models,
such as the values of the transport coefficients and, crucially, their dependence on
hydrodynamic fields. As a result, they typically miss the density-segregated regime and its nonlinear structures.

Other approaches, rooted in a direct coarse-graining of the
microscopic dynamics \cite{IGOR-LEV,BDG,MCM-rod1,MCM-polar,MCM-rod2,IHLE,PAWEL}, on the other hand, are able to determine
most transport coefficients and their functional dependence on the fields.
The ``Boltzmann-Ginzburg-Landau" (BGL) framework, which we detail below,  
stems from the early Boltzmann approach of \cite{BDG}
to the polar case, and combines it with the traditional Ginzburg-Landau weakly nonlinear analysis.
We argue below that it offers better overall control, something needed since  
some confusion remains: the equations obtained by different methods often differ not only in details
but also in structure. 
For instance, the equations for rods derived by Baskaran and Marchetti from a Smoluchowski equation contain
more and different terms than those derived in the BGL framework. The Chapman-Enskog formalism put forward
in \cite{IHLE} for the polar case yields many more terms than those retained in other approaches, and their 
effect on the dynamics, especially at the nonlinear level, remains unclear. 
 
Recent work has shown, we believe, that the 
Boltzmann-Ginzburg-Landau framework is near-ideal for deriving minimal, well-behaved, nonlinear hydrodynamic descriptions
from Vicsek-like models, without aiming at a quantitative agreement with what remains, after all, rather unrealistic starting points.
The equations derived for the three basic classes however have, we believe, universal value and 
have been shown to account remarkably well for most features observed, 
including highly nonlinear solutions and chaotic regimes.

In the following, we present a unified formulation of the BGL
approach which is applicable to generic dry active matter systems in their dilute limit and
encompasses both ballistic propagation and positional (anisotropic)
diffusion. This allows one to treat  both {\it propagative} (polar case, rods) and 
{\it diffusive} (active nematics) active systems, as well as systems where
non-equilibrium activity and ``thermal" (random) agitation compete.

In Section~\ref{S2}, we define the generalized Vicsek model we start from and derive the associated Boltzmann
kinetic equation. In Section~\ref{S3}, we derive hydrodynamic equations for the three basic classes. 
In Section~\ref{S4}, we discuss the BGL framework, the obtained results, and compare them to other works.

\section{Vicsek-like models and their Boltzmann description} 
\label{S2}

In dry systems, motion takes place over a substrate or in a fluid which acts as a momentum sink, so that overall (active
particles plus fluid) momentum conservation is not a concern and long-ranged hydrodynamic interactions between
particles can be neglected (hence the term dry). Over the elementary dissipative timescale $\Delta t$ of the dynamics, particle motion is coupled to particle orientation, leading either to self-propulsion or to active diffusion.
From a modeling viewpoint, active particles are represented here by point-like objects interacting locally through alignment `forces'
in the presence of noise.
The alignment and the noise are a simple means to model the (possibly) complicated interactions between active particles in real systems.
Due to the strongly dissipative nature of the substrate/surrounding fluid, our point particles follow an effective first order dynamics, both positionally and rotationally.

\subsection{Definition of a generalized Vicsek model}

We consider active particles moving in a two dimensional
continuous space. Particles are described by their position ${\bf r}$,
and a (unit) heading vector defined by an
angle $\theta \in [-\pi,\pi]$. \\
In a generalization of the classical VM
streaming rule, particles positions are updated according to 
\begin{equation}
{\bf r}(t+\Delta t) ={\bf r}(t)+ \, {\bf v} \, \Delta t
\label{streaming}
\end{equation}
where ${\bf v}$ is a random vector drawn from a displacement distribution
$\Phi(v,\theta_v-\theta)$, with $v=|{\bf v}|$ and $\theta_v$
being the angle defining the orientation of ${\bf v}$. In the absence
of chirality one has $\Phi(v,\theta_v-\theta)=\Phi(v,\theta-\theta_v)$.
Purely ballistic motion with speed $v_0$ (standard VM dynamics), for instance, 
is recovered considering the delta distributions
$\Phi(v,\theta_v-\theta) = \delta(v-v_0)
\delta(\theta_v-\theta)$.
The angular part of the distribution
$\Phi(v,\theta_v-\theta)$  is obviously $2\pi$-periodic, but higher
order symmetries may reflect further symmetries in particles
displacement. Active nematic particles, for instance, are
characterized by $\pi$-periodicity (and of course no higher order symmetries). 
For compactness, we shall say that $\Phi$ is $m\pi$-periodic, with
$m=1$ in the nematic case and $m=2$ in the polar
case.

The heading angle $\theta$ of a given particle evolves according to the simple stochastic
dynamics,
\begin{equation}
\theta' =
\Psi^{(p)}(\theta,\theta_{i_1},\theta_{i_2},\ldots,\theta_{i_p}) +
\eta \;,
\label{alignment}
\end{equation}
where $\eta$ is a symmetrically-distributed (zero mean) random angle (delta
correlated in time and between different particles) and
$\theta_{i_1},\theta_{i_2},\ldots,\theta_{i_p}$ are the angles of the
$p$ interacting particles which, in {\it metric models}, are typically
defined as the particles closer than the metric interaction range $r_0$. When no
neighbours are present, particles  simply experience self-diffusion events
$\Psi^{(0)}(\theta)=\theta+\eta$.

Contrary to the orientational degree of freedom, the modulus of the displacement vector {\bf v} has essentially no dynamics, 
and is assumed to be a hard mode\footnote{However, recent experimental results
\cite{SCALE-FREE} suggest this may not be the case in starling flocks.}
with only fast fluctuations around its constant mean value.


\subsection{Boltzmann equation}

The above generic model can be described in the standard framework of
kinetic theory, through the introduction of a (generalized) Boltzmann
equation, which describes the time evolution of the
one-particle distribution function $f({\bf r}, \theta, t)$, namely the probability to find a particle at position ${\bf r}$ at time $t$. We choose the normalization 
\begin{equation}
\frac{1}{V} \int_V d{\bf r} \int_{-\pi}^{\pi} d\theta f({\bf r}, \theta, t)=\rho_0
\end{equation}
with $V$ the system volume and $\rho_0$ its mean density.

The Boltzmann approach
relies on the {\it molecular chaos} hypothesis and the {\it binary
collision approximation}, which, to a certain extent, are justifiable
in low density systems. We will comment on these approximations in
section \ref{S4}.


\subsubsection{Positional part}

To obtain the Boltzmann equation, we first consider the (generalized) ``streaming''
part of the dynamics, described by Eq.~(\ref{streaming}).
One has
\begin{equation} \label{Eq-Boltz-stream0}
f({\bf r}, \theta, t+\Delta t)= \int d{\bf v}\, \Phi(v,\theta_v-\theta)
f({\bf r} - {\bf v} \Delta t, \theta, t).
\end{equation}
On time scales much larger than $\Delta t$, applying It\^o calculus
\cite{Ito} to second order, one obtains the master equation

\begin{equation}
\frac{\partial f}{\partial t} =- \langle v_\alpha
\rangle \partial_\alpha f({\bf r}, \theta, t) 
 + \frac{1}{2} \langle \delta v_\alpha \delta v_\beta \rangle \Delta t \,
\partial_\alpha \partial_\beta f({\bf r}, \theta, t) 
\end{equation}
where summation over the repeated spatial indices $\alpha$, $\beta$ is
understood and the brackets $\langle \ldots \rangle$ denote an average over the distribution $\Phi(v,\theta_v-\theta)$
\begin{eqnarray}
\langle v_\alpha \rangle &=& \int_0^{\infty} dv \int_{-\pi}^{\pi}
d\theta_v\, v n_{\alpha}(\theta_v)\, \Phi(v,\theta_v-\theta)\\
\langle \delta v_\alpha \delta v_\beta \rangle &\!=\!& \!\!\int_0^{\infty} \!\!\!dv \!\int_{-\pi}^{\pi}
\!\!d\theta_v\, \delta v_{\alpha}(\theta_v) \delta v_{\beta}(\theta_v) \Phi(v,\theta_v\!-\!\theta)
\end{eqnarray}
with ${\bf n}(\theta)=(\cos \theta, \sin \theta)^T$ the unit vector in the direction $\theta$,
${\bf v}=v {\bf n}(\theta)$ and $\delta v_{\alpha}=v_\alpha - \langle v_\alpha \rangle$.

For polar particles, the vector
$\langle {\bf v} \rangle$ points in the direction defined by $\theta$.
One can then write $\langle {\bf v} \rangle = v_0 {\bf n}(\theta)$,
which defines the average speed of particles $v_0=\langle
v\cos\delta\theta\rangle$ ($\delta\theta \equiv \theta_v-\theta$). 
For nematic particles $\langle {\bf v} \rangle=0$, and one can thus also formally write $\langle {\bf v} \rangle = v_0 {\bf n}(\theta)$, with $v_0=0$.
The covariance term $\langle \delta v_\alpha \delta v_\beta \rangle$ is less straightforward to evaluate.
After some algebra 
we find
\begin{eqnarray}
\langle \delta v_\alpha \delta v_\beta \rangle = \frac{1}{2} \langle
v^2 (1-\cos2\delta\theta)\rangle \delta_{\alpha \beta}
+ \left(\langle v^2 \cos2\delta\theta \rangle -v_0^2\right)
n_\alpha(\theta) n_\beta(\theta)
\end{eqnarray}
which can be recast as the sum of an isotropic ($\sim
\delta_{\alpha\beta}$) and an anisotropic ($\sim g_{\alpha \beta} =
n_{\alpha}n_{\beta} - \delta_{\alpha\beta}/2$) term. Here $
\delta_{\alpha\beta}$ is a Kronecker delta and $g_{\alpha\beta}$ is
the nematic tensor familiar to liquid crystal physics \cite{LiquidC}.
We write the positional part of the Boltzmann equation as
\begin{equation}
\frac{\partial f}{\partial t} = - v_0
n_{\alpha} \partial_{\alpha}f + D_0 \Delta f
+ D_1 \,g_{\alpha\beta}\, \partial_\alpha \partial_\beta f
\end{equation}
where $\Delta f$ is the Laplacian of $f$ and
\begin{eqnarray}
D_0&=& \frac{1}{4} \left(\langle v^2\rangle -v_0^2\right) \Delta t\\
D_1&=&\frac{1}{2} \left( \langle v^2 \cos2\delta\theta \rangle -v_0^2 \right) \Delta t
\end{eqnarray}
are, respectively, the isotropic and anisotropic diffusion constants.

Fluctuations in the displacement vector ${\bf v}$, either in modulus or
in heading, lead to a finite isotropic diffusion term
$D_0>0$. Anisotropic diffusion ($D_1 \neq 0$), on the other hand, is
produced by anisotropies in the fluctuations of the displacement
vector ${\bf v}$.
In this regard, it is instructive to consider the simple ``4
directions'' case
\begin{equation}
\Phi(v,\delta \theta) = \delta (v-v_s)\sum_{j=1}^4\ p_j \, \delta
\left(\delta \theta - (j-1)\frac{\pi}{2}\right) 
\end{equation}
where $v_s$ is the speed of the particles,
$p_j\geq0$, and $\sum_j p_j=1$. For illustrative
reasons, we can temporarily relax the non-chiral condition and allow $p_2\neq p_4$. 
One then has a drift term with mean velocity $\langle {\bf v} \rangle = v_0 {\bf n} (\theta + \psi_0)$
where 
\beq
v_0 = v_s \sqrt{(p_1-p_3)^2 +(p_2-p_4)^2}
\eeq
and
\beq
\psi_0 = \mathrm{Arg}\left[ (p_1-p_3) + i (p_2-p_4)\right]
\eeq 
($i$ being the imaginary unit).

Diffusion constants can be written as $D_0=(D_\parallel + D_\perp)/2$ and
$D_1=(D_\parallel - D_\perp)$, where 
\begin{eqnarray}
D_\parallel&=&\left[p_1+p_3-(p_1-p_3)^2\right]\frac{v_s^2}{2} \Delta t \\
\label{eq:D1}
D_\perp&=&\left[p_2+p_4-(p_2-p_4)^2\right]\frac{v_s^2}{2} \Delta t \nonumber
\label{eq:D2}
\end{eqnarray}
highlighting the proportionality of $D_1$ to the difference between 
parallel and perpendicular fluctuations (w.r.t. local heading).

\subsubsection{Full Boltzmann equation}

In the low density limit, interactions between more than two particles
are unlikely, and the heading angle dynamics is captured by self-diffusion events and binary collision-like events.
In self-diffusion events, the heading angle $\theta$ is changed by a random amount $\eta$, 
drawn from a zero-mean, symmetric distribution $P_\sigma(\eta)$ with standard
deviation $\sigma$.  As a slight simplification, we assume that such self-diffusion events
occur with a constant rate $\lambda$ (typically of the order of $1/\Delta t$), independent of the density.
In binary collisions, the incoming angles $\theta_1$ and $\theta_2$ of the two particles are changed, after collision, into
\begin{equation} \label{angle-update}
\theta_1' = \Psi(\theta_1,\theta_2) + \eta_1 \;,
\quad
\theta_2' = \Psi(\theta_2,\theta_1) + \eta_2
\end{equation}
where $\eta_1$ and $\eta_2$ are random variables drawn from
 $P_\sigma(\eta)$. In principle, the collision and self-diffusion noise distributions
could be different. To simplify the presentation, we shall consider
in what follows that the two distributions are the same; however 
calculations can be easily carried out with distinct
distributions.

The collision rate is encoded in a collision kernel
$K(\Delta=\theta_2-\theta_1) \ge 0$, which depends only on the angle
difference due to a global rotational invariance of the
problem. In metric models, where the collision probability can be
modeled as a scattering process \cite{KH}, the collision kernel $K(\Delta)$ 
inherits the same $m\pi$-symmetries as that of the displacement distribution $\Phi$.

Interaction symmetries also impose constraints on the interaction rule
$\Psi(\theta_1,\theta_2)$, which can typically be either polar or nematic depending on its periodicity with respect to each angle $\theta_1$ and $\theta_2$.
Formally, $\Psi(\theta_1,\theta_2)$ is $n\pi$-periodic, modulo $n\pi$, with
respect to both arguments $\theta_1$ and $\theta_2$ independently, which defines the symmetry index $n$ of the interaction rule.
In the following, we call ``ferromagnetic'' such an alignment rule with
$2\pi$-periodicity ($n=2$) and nematic an alignment rule with
$\pi$-periodicity ($n=1$). The interaction rule must also be
consistent with the displacement symmetry of particles (reflected in
the periodicity of the displacement distribution $\Phi$ and of kernel $K$), which implies $n \le m$.

In addition, isotropy imposes that for an arbitrary rotation of angle $\phi$,
\begin{equation}
\Psi(\theta_1+\phi,\theta_2+\phi) = \Psi(\theta_1,\theta_2) + \phi \; [n\pi].
\end{equation}
Choosing $\phi=-\theta_1$ and $\theta_2=\theta_1+\Delta$, one obtains
\begin{equation}
\Psi(\theta_1,\theta_1+\Delta) = \theta_1 + \Psi(0,\Delta) \; [n\pi].
\end{equation}
Hence the interaction rule $\Psi(\theta_1,\theta_2)$ is parameterized
by a single variable interaction function $H(\Delta) \equiv \Psi(0,\Delta)$, which is $n\pi$-periodic.
We further assume that there is no chirality in the problem and that
the particle exchange symmetry is respected by point particles
collisions, implying that $K(-\Delta)=K(\Delta)$ and
$H(-\Delta)=-H(\Delta)\, [n\pi]$. Due to these symmetry properties, we need only define $K(\Delta)$ and $H(\Delta)$ over the positive half of their definition interval.

The full Boltzmann equation is obtained by taking into account the self-diffusion and binary collision rules (assumed to be valid in the low density regime considered here),\footnote{Note that strictly speaking, the Boltzmann equation describes a version of the microscopic model in which interactions takes place only once during a collision, when the interparticle distance reaches the interaction range. No further interaction occurs during the phase when particles are closer than the interaction range. A new interaction takes place when particles collide again.}
under the molecular chaos assumption 
which approximates the two-particle distribution $f^{(2)}$ as the product
of two one-particle distributions
\begin{equation}
f^{(2)}({\bf r},\theta_1, \theta_2, t) \approx f({\bf
  r},\theta_1, t) f({\bf r},\theta_2, t) .
\end{equation}
Physically, this assumption means that the heading angle decorrelates between consecutive collisions.
One then finds for the Boltzmann equation
\begin{equation}
\frac{\partial f}{\partial t} +v_0
n_{\alpha} \partial_{\alpha}f = D_0 \Delta f
+ D_1 \, g_{\alpha \beta} \,\partial_\alpha \partial_\beta f
+ I_{\rm dif}[f] + I_{\rm col}[f] 
\label{FullBoltzmann}
\end{equation}
where the self-diffusion integral $I_{\rm dif}[f]$ and the collision integral $I_{\rm col}[f]$ are defined as
\begin{eqnarray}
\label{Idif}
I_{\mathrm{dif}}[f] &=& -\lambda f(\mathbf{r},\theta,t)
+ \lambda \int_{-\pi}^{\pi} d\theta' \int_{-\infty}^{\infty} d\eta\,
P_{\sigma}(\eta)
\delta_{m\pi}(\theta'+\eta-\theta) f(\mathbf{r},\theta',t)\\
\label{Icol}
I_{\mathrm{col}}[f] &=& - f(\mathbf{r},\theta,t)
\int_{-\pi}^{\pi} d\theta'\,
K(\theta'-\theta)\, f(\mathbf{r},\theta',t)\\
\nonumber
&& + \int_{-\pi}^{\pi} d\theta_1 \int_{-\pi}^{\pi} d\theta_2
\int_{-\infty}^{\infty} d\eta\, P_{\sigma}(\eta)\, K(\theta_2-\theta_1)
f(\mathbf{r},\theta_1,t) f(\mathbf{r},\theta_2,t)\\
\nonumber
&& \qquad \qquad \qquad \qquad \qquad \qquad \qquad \qquad \times \, \delta_{m\pi}\Big(\Psi(\theta_1,\theta_2)+\eta-\theta\Big)
\end{eqnarray}
where $\delta_{m\pi}$ is a generalized Dirac delta imposing that the argument is equal to zero modulo $m\pi$.

Without any loss of generality, we can set $\lambda=1$ (this amounts to
a rescaling of time).
For convenience, we further consider the noise distribution $P_\sigma$ to
be Gaussian. 
The physics described by the Boltzmann equation is determined by just three functions encoding
 the microscopic dynamics: the displacement distribution $\Phi$,
the collision kernel $K$ and the interaction function $H$.

\subsection{Three basic classes}

\label{sec-def-class}

A trivial case for the Boltzmann equation (\ref{FullBoltzmann}) is
given by a displacement distribution $\Phi$ with an isotropic angular
part, for which one has $v_0=0$, $D_0\sim d_0^2/4$ and $D_1=0$.
This corresponds to an isotropic diffusive motion, which can be mapped
onto a passive, equilibrium system. Here displacement is strictly random
and completely decoupled from orientational dynamics.

Anisotropies in $\Phi$, on the other hand, lead to non-equilibrium activity.
Although this does not cover all possible cases, it is relatively easy
to identify the three simplest classes depending on the symmetries of the
particles displacement ($m\pi$-periodicity of the displacement
distribution $\Phi$) and of the interactions ($n\pi$-periodicity of
the interaction rule $\Psi$). These three classes arise from the fact
that the symmetry (polar or nematic) of both the particles and the
interaction rule can be varied, with however the constraint that the interaction rule must obey at least the particle symmetry ($n \le m$).


\subsubsection{Polar particles with ferromagnetic interaction}
This class, labeled as $m=n=2$, is exemplified by
the Vicsek model for self-propelled particles.
Displacement distributions breaking the nematic symmetry $\theta_v \to
\theta_v +\pi$ result in a nonzero drift, $v_0 \neq 0$. Such active
particles are typically described as {\it self-propelled}. In metric systems
dominated by drift, $K(\Delta)$ is $2\pi$-periodic (polar
symmetry); for an isotropic interaction range, it is given by the flux of incoming particles through the
cross-section $2 r_0$ of a target particle \cite{BDG}, 
\begin{equation}
K(\Delta)=2 v_0\,r_0 \left| {\bf n}(\theta_2)-{\bf n}(\theta_1)\right
|=4 v_0\,r_0\left|\sin \frac{\Delta}{2}\right| .
\label{pkernel}
\end{equation}
However, more complex forms of $K(\Delta)$ (typically involving higher
harmonics in $\Delta$) can be used to describe
anisotropic interaction ranges (as the one used to model collisions of
elongated objects in a point-like framework \cite{Weber2013}).

The interaction function $H(\Delta)$ describing ferromagnetic interactions is $2\pi$-periodic.
For the ``canonical'' Vicsek model studied in \cite{VICSEK,CHATE}, the function $H(\Delta)$
is given by $H(\Delta)=\Delta/2$ 
and diffusion terms strictly vanish, with $\Phi(v, \delta \theta) =
\delta(v-v_0)\delta_{2 \pi}(\delta \theta)$ implying $D_0=D_1=0$.
If the displacement distribution is characterized by a finite
variance, non-zero diffusion is added to drift. Isotropic diffusion is
characterized by $D_0>0$ and can be interpreted as having a thermal
origin. The anisotropic part of diffusion, on the other hand, arises from
fluctuation anisotropies and has (in the present framework of point-like particles) a strictly non-equilibrium origin \cite{GNF1}. Indeed, speed fluctuations in the presence of drift 
are enough to generate anisotropic
diffusion even in the absence of heading fluctuations.

\subsubsection{Polar particles with nematic interaction}
This class, labeled as $m=2$, $n=1$, corresponds to self-propelled
($v_0 \neq 0$) ``rods", namely self-propelled particles with an elongated shape making their alignment
nematic.
The kernel $K(\Delta)$ is $2\pi$-periodic (to zeroth order in particle anisotropy), and for metric systems is still given by Eq.~(\ref{pkernel}), 
but $H(\Delta)$ is now $\pi$-periodic.
For the representative model of this class, studied in \cite{RODS,RODS-KINETIC}, one has $H(\Delta)=\frac{\Delta}{2}$, and $\Phi(v, \delta \theta) =
\delta(v-v_0)\delta_{2 \pi}(\delta \theta)$.

\subsubsection{Nematic particles with nematic interaction}

This case $m=n=1$ correspond to the ``active nematics''
class. Here the displacement distribution is $\pi$-symmetric and
drift is strictly zero, with the system dominated by finite isotropic and
anisotropic (non-equilibrium) diffusion.
Both $K(\Delta)$ and $H(\Delta)$ are $\pi$-periodic.
In the simple case of isotropic interaction range, the polar kernel
(\ref{pkernel}) can be modified to account for nematic symmetry
\begin{equation}
K(\Delta)=2 v_0\,r_0 \left|\sin \frac{\Delta}{2}+\cos \frac{\Delta}{2}\right|\,.
\label{nkernel}
\end{equation}
A representative model of this class studied in the literature \cite{Chate2006,ACTIVE-NEMA}
is defined by $H(\Delta)=\frac{\Delta}{2}$ and $\Phi(v, \delta \theta) = \delta(v-v_0) \delta_{\pi}(\theta)$.

\subsection{Metric-free models}
\label{metric-free}

So far, we have mostly discussed models characterized by metric interaction ranges.
Motivated by recent results on the observation of starling flocks
\cite{Ballerini}, and fish schools \cite{Gautrais}, models where the interaction range is not defined by a metric
distance, but rather by a typical number of neighbors, are of interest.
Metric-free versions of the Vicsek model, where neighbours are not chosen inside a
metric range but rather by some topological criteria (for instance,
making use of Voronoi tessellations) have been introduced in
\cite{TOPOVICSEK}. From a theoretical perspective, a difficulty with
such models is that a particle
always interacts with its neighbors, whatever their
metric distance. In order to make such a situation suitable for a kinetic
theory approach, it has been proposed to introduce a low interaction
rate such that at each time step, two neighboring particles (in a
topological sense) have only a small probability to interact
\cite{TOPOKINETIC}. 

Once two particles have interacted, their velocity angles are updated according to Eq.~(\ref{angle-update}).
The difference with the metric cases appears in the
collision kernel $K$. Here interactions cannot be described by a
scattering process, and the probability of collision does not depend on the angles of the two
interacting particles. 
In metric models, $K$ is defined under the implicit assumption
that the probability of collision is proportional to the local
particles density $\rho({\bf r}, t) \equiv
\int_{-\pi}^\pi d\theta f({\bf r}, \theta, t)$. 
On the other hand, for metric-free models, the probability of
interaction is independent of the local density and to take this crucial
effect into account, one uses the framework of metric models, 
but with an effective interaction kernel proportional to $1/\rho$.

\section{Hydrodynamic equations} 
\label{S3}

In order to derive hydrodynamic equations from the Boltzmann equation, we proceed as follows.
First, we expand the Boltzmann equation in angular Fourier
modes, yielding an infinite hierarchy of equations for these modes.
Second, we look for the first linearly unstable angular mode when decreasing the noise and increasing the density, and truncate the hierarchy of equations in the vicinity of this linear instability.
The continuous description thus includes the (conserved) local density field and the relevant angular modes (typically the linearly unstable mode).

The specificity of the "Boltzmann-Ginzburg-Landau" approach is to rely on a
systematic scaling ansatz to truncate and close the hierarchy to a set of minimal equations.

\subsection{Angular Fourier modes expansion}

Here we chose to restrict ourselves to two spatial dimensions, where the use of Fourier
transforms and complex notation greatly simplifies calculations.
In order to define a framework consistent with the most general
$2\pi$-symmetry, we work with an angular Fourier expansion defined over the $2\pi$ interval:
\begin{equation} \label{Fourier-exp}
f(\mathbf{r},\theta,t) = \frac{1}{2\pi} \sum_{k=-\infty}^{\infty}
\hat{f}_k(\mathbf{r},t) \, e^{-ik\theta},
\end{equation}
where the Fourier coefficients $\hat{f}_k$ are defined as
\begin{equation}
\hat{f}_k(\mathbf{r},t) = \int_{-\pi}^{\pi} d\theta \,
f(\mathbf{r},\theta,t)\, e^{ik\theta}.
\end{equation}

Note that $\hat{f}_0(\mathbf{r},t)$ is nothing but the local density $\rho(\mathbf{r},t)$. Note also
that for all $k$, $\hat{f}_{-k}=\hat{f}_{k}^*$ (the star denotes the complex conjugate).
An additional $\pi$-periocity of the one particle distribution $f$ then manifests itself by the 
nullity of odd Fourier coefficients, $\hat{f}_{2l+1}=0$. 
In the following, we drop the 'hat' over the Fourier coefficients in order to lighten notations.

It is easy to verify that the Fourier modes $f_k$ ($k \ge 1$) can be
interpreted as order parameter fields associated to specific spontaneous symmetry
breakings. In particular, $f_1$ encodes the momentum field, and $f_2$ the nematic tensorial field:
\begin{equation}
\rho {\bf P} \!=\! \left(\!\begin{array}{lr}
{\rm Re}f_1\\
{\rm Im}f_1 \end{array} \!\right) , \;
\rho {\bf Q} \!=\! \frac{1}{2} \!\left(\begin{array}{lr}
{\rm Re}f_2 & {\rm Im}f_2 \\
{\rm Im}f_2 & -{\rm Re}f_2\end{array} \!\right) \, ,
\end{equation}
where ${\bf P}$ is a polarity field of components $P_\alpha= \langle n_\alpha \rangle_l$, 
and ${\bf Q}$ the traceless tensorial field of components
$Q_{\alpha\beta}=\langle  n_\alpha n_\beta \rangle_l - \delta_{\alpha\beta}/2$
(here $\langle \ldots \rangle_l$ denotes a local average over the distribution $f$.)

Introducing the complex derivatives $\nabla = \partial_x + i\partial_y$ and
$\nabla^* = \partial_x - i\partial_y$ as well as the Laplacian
$\Delta=\nabla\nabla^*$ (note that in this notation $\nabla^2=(\partial_x + i\partial_y)^2$ is not
the Laplacian), the angular Fourier expansion of the
Boltzmann equation then yields
the infinite hierarchy of coupled equations
\begin{eqnarray} \nonumber
\frac{\partial{{f}_k}}{\partial t} + \frac{v_0}{2} ( \nabla {f}_{k-1} + \nabla^* {f}_{k+1} ) &=& - (1-{P}_k) {f}_k + D_0 \Delta {f}_k
+ \frac{D_1}{4} (\nabla^2 {f}_{k-2}+\nabla^{*2} {f}_{k+2})\\
&& \qquad \qquad \qquad + \sum_{q=-\infty}^{\infty} ({P}_k I_{k,q}-I_{0,q}) {f}_q {f}_{k-q}
\label{eq-fk}
\end{eqnarray}
where the coefficient $I_{k,q}$ is defined by the integral
\begin{equation}
I_{k,q} = \frac{1}{2\pi} \int_{-\pi}^{\pi} d\Delta \, K(\Delta)\, e^{-iq\Delta+ikH(\Delta)}
\end{equation}
and ${P}_k \equiv {P}_k(\sigma) = \int_{-\infty}^{\infty} d\eta\, P_\sigma(\eta)\, e^{ik\eta}$ is the Fourier transform of the noise distribution
(restricted here to integer values of $k$). One has $0 \leq
{P}_k(\sigma) \leq 1$ and ${P}_k(0)=1$ $\forall k$.
For a Gaussian noise distribution, the Fourier transform has the simple form
\begin{equation}
{P}_k(\sigma)=e^{-\sigma^2 k^2/2} .
\end{equation}
Note also that due to the parity properties of $K(\Delta)$ and $H(\Delta)$, $I_{k,q}$ is real.

From Eq.~(\ref{eq-fk}), we find in particular the continuity equation, obtained for $k=0$
\begin{equation} \label{eq-continuity}
\frac{\partial \rho}{\partial t} + v_0 {\rm Re} (\nabla^* f_1) = 
D_0 \Delta \rho
+ \frac{D_1}{2} {\rm Re} (\nabla^{*2} f_2).
\end{equation}
Note that this equation is valid without any further assumption, due
to the fact that the Fourier transform of the integral terms in Eq.~(\ref{FullBoltzmann}) vanish for $k=0$.

\subsection{Linear instability of the disordered state}

\subsubsection{General considerations}

The homogeneous disordered isotropic state $f_0=\rho_0$, $f_k=0$ ($k \ge 1$) is a trivial
solution of Eq.~(\ref{eq-fk}). Assuming spatial homogeneity, we
linearize Eq.~(\ref{eq-fk}) around the isotropic solution, yielding for $k>0$
\begin{equation}
\frac{\partial{f_k}}{\partial t} = \left[ -(1-{P}_k) 
+\omega_k \rho_0 \right] f_k
\end{equation}
where we have defined
\begin{equation}
\omega_k = P_k (I_{k,k}+I_{k,0}) - (I_{0,k}+I_{0,0}).
\end{equation}
The linear stability of $f_k$ is governed by the sign of the linear
coefficient, 
\begin{equation} \label{def-muk}
\mu_k (\sigma, \rho_0) \equiv -(1-{P}_k) +\omega_k \rho_0.
\end{equation}
Analyzing the full noise and density dependence of this expression for
arbitrary kernel and interaction rule is a complicated
task. One can however notice that at low enough noise ${P}_k \to
1$ and the sign of
$\mu_k$ is then given by that of $\omega_k$, since $\mu_k \approx \omega_k
\rho_0$ in this limit. By studying $\omega_k$ instead of $\mu_k$, one
thus gets rid of the $\rho_0$ dependence.
Therefore, the sign of $\omega_k$ is controled by the zero noise limit $\omega_k^{(0)}=(I_{k,k}+I_{k,0}) - (I_{0,k}+I_{0,0})$: if $\omega_k^{(0)} \ne 0$, there exists a finite range of $\sigma$, in the vicinity of $\sigma=0$, where $\omega_k$ has the same sign as $\omega_k^{(0)}$. Hence, the study of the quantity $\omega_k^{(0)}$, which depends only on the kernel and on the interaction rule (and not on the noise and density) should be enough to detect an instability of $f_k$. Of course, this simplification imposes several limitations. First, by construction, the study of $\omega_k^{(0)}$ does not allow in itself for the identification of an instability line in the parameter plane ($\sigma$, $\rho_0$). Second, it provides no information on which mode is the most unstable close to the instability threshold in case several modes would be unstable in the high density and low noise limit. However, if a single mode is found to be unstable in this limit, one is sure that this mode should be taken as the order parameter of the transition.

An explicit expression of $\omega_k^{(0)}$ is given by
\begin{equation}
\omega_k^{(0)} = \frac{1}{\pi} \int_0^{\pi} d\Delta K(\Delta)
\Big[ \cos k(H(\Delta)-\Delta) + \cos kH(\Delta) - \cos k\Delta -1 \Big].
\end{equation}
For any given specific kernel and interaction rule, one can compute
(analytically or numerically) the coefficients $\omega_k^{(0)}$ for
$k=1,\,2,\dots$, and evaluate their sign to
determine the first unstable mode. In particular, note that in
metric-free models, due to the dependence of the effective kernel on
density one has $\omega_k^{(0)}\sim 1/\rho$, resulting in a linear
coefficient $\mu_k (\sigma)$ independent of density.

\subsubsection{First unstable mode: study of generic classes}
\label{Sec:alignment}

In the following, we try to determine the sign of $\omega_1^{(0)}$ and $\omega_2^{(0)}$ as a function of the symmetries of the kernel $K(\Delta)$ and the interaction rule $H(\Delta)$.
The fully general case is hard to analyze, but we are able to obtain fairly general results for each of 
the three classes introduced in Sect.~\ref{sec-def-class}, under the additional assumption that the interaction rule is indeed an alignment rule for all values of the angle difference $\Delta$, which translates in mathematical terms as $0 \le H(\Delta) \le \Delta$ for $0 < \Delta < \frac{n\pi}{2}$,
with $0 < H(\Delta) < \Delta$ over at least a finite subinterval of $\Delta$.
The main results can be summarized as follows.

\begin{itemize}

\item {\it Polar particles with ferromagnetic alignment}

One generically finds that $\omega_1^{(0)}>0$, while the sign of $\omega_2^{(0)}$ cannot be determined without further assumptions on the interaction rule.
If $H(\Delta)$ is `close enough' to the standard rule $H(\Delta)=\frac{\Delta}{2}$, one recovers $\omega_2^{(0)}<0$. It is likely that close to the transition line $\omega_1=0$ in the noise-density plane, one has $\omega_2<0$ so that only polar order is unstable, but checking this for arbitrary rules within this class is a difficult task.
In any case, this result shows that any generic ferromagnetic aligning
rule which respects the particle exchange symmetry
$H(-\Delta)=-H(\Delta)$ is enough to make the disordered solution
unstable towards the growth of polar order.\\

\item {\it Polar particles with nematic alignment}

This case somehow mirrors the previous one, exchanging the roles of
$\omega_1^{(0)}$ and $\omega_2^{(0)}$. Under the generic assumptions
of this class, one finds that $\omega_2^{(0)}>0$, pointing to a
nematic instability for generic nematic alignment rules. The sign of
$\omega_1^{(0)}$ is not completely fixed within the class. However,
under the fairly natural assumption that $K(\Delta)$ is a
non-decreasing function in $[0, \frac{\pi}{2}]$, as is the case of the
kernel (\ref{pkernel}), one can show that $\omega_1^{(0)}<0$. This result remains true if $K(\Delta)$ decreases only slightly.\\

\item {\it Nematic particles with nematic alignment}

Similarly to the case of polar particles with nematic interaction, one
finds $\omega_2^{(0)}>0$ for generic nematic alignment rules.
The situation is however simpler here, since $f_1=0$ by symmetry, so that its dynamics need not be studied.

\end{itemize}

Let us emphasize that the 'contracting' condition on $H(\Delta)$, namely $0 \le H(\Delta) \le \Delta$,
is only a sufficient condition allowing for easier calculations, but instabilities of the disordered state can also occur if $H(\Delta)$ is not 'contracting' everywhere.
As an illustrative example of a polar rule that is not 'contracting' everywhere, let us consider the function $H(\Delta)$ defined as $H(\Delta)=2\Delta$ for $0<\Delta<a\pi$ and $H(\Delta)=\frac{\Delta}{2}$ for $a\pi<\Delta<\pi$, with $0<a<1$.
One then finds for $\omega_1^{(0)}$, assuming for simplicity that $K(\Delta)$ is a constant $K_0$, independent of $\Delta$,
\begin{equation}
\omega_1^{(0)} = \frac{K_0}{\pi} \left( 4-\pi + \frac{1}{2} \sin 2a\pi + \sin a\pi -4\sin \frac{a\pi}{2} \right)
\end{equation}
which reduces for small $a$ to
\begin{equation}
\omega_1^{(0)} \approx \frac{K_0}{\pi} \left( 4-\pi - \frac{3}{4}  a^3 \pi^3 \right).
\end{equation}
For small enough  $a$, $\omega_1^{(0)}$ is thus positive, and the polar order parameter $f_1$ is unstable at low noise and/or high density.

\subsection{Scaling ansatz close to instability threshold}

Our goal is to reduce the infinite hierarchy (\ref{eq-fk}) of equations to a small set of minimal equations for the relevant hydrodynamic fields. The number of particles being conserved, the density $\rho({\bf r},t)$ is a relevant field, for which we already have derived the evolution equation (\ref{eq-continuity}).
Additional relevant fields correspond to broken symmetries, and can be
identified from the linear instabilities studied above. In order to
obtain a finite number of equations, one needs to truncate the
infinite hierarchy of equations, which requires some truncation
criterion. In the spirit of Ginzburg-Landau equations, we shall use a
scaling ansatz close to the instability threshold of the first
unstable mode in which the order parameter fields are small quantities
and, being interested in the long-wavelength long-timescale
dynamics, also time and space derivatives of the fields are smaller
than the field themselves. 

In the following, we shall first review how to obtain this scaling ansatz
in each of the three basic classes outlined above, and next we will
use what we have learned to formulate a systematic scaling ansatz valid
in the general case. 
The simplest case is that of active nematics, so we start by considering this case.

\subsubsection{Nematic particles with nematic interaction}
\label{sec-active-nema-scal}

The only relevant order parameter here is $f_2$.
We assume that close to the instability threshold, $f_2$ is of the order of a small parameter $\epsilon$, related to the small value of the linear coefficient $\mu_2>0$.
The continuity equation (\ref{eq-continuity}), here with $f_1=0$,
imposes two constraints: (i) density fluctuations $\delta
\rho=\rho-\rho_0$ with respect to the average density $\rho_0$ are of
order $\epsilon$; (ii) if spatial derivatives $\nabla$ are of order
$\epsilon^{\alpha}$, then the time derivative $\partial_t$ scales as
$\epsilon^{2\alpha}$. Nonlinear saturation should result from the
interaction between $f_2$ and the next non zero mode $f_4$. Assuming
that the coefficient $\mu_4$ is negative, $f_4$ should be slaved to
$f_2$ in the equation for $f_4$ since $|\partial_t f_4| \ll |\mu_4
f_4|$. Hence the linear term in $f_4$ should be balanced by the term in $f_2^2$ in the equation for $f_4$, yielding $f_4 \sim \epsilon^2$.

Coming back to the equation for $f_2$, one has to balance (again in a
Ginzburg-Landau spirit) the diffusive term $\Delta f_2$ with the
nonlinear term $f_2^* f_4$, which fixes $\alpha=1$. In summary, one has:
\begin{equation} \label{eq-active-nema-scal}
f_2 \sim \epsilon, \quad f_4 \sim \epsilon^2, \quad \rho-\rho_0 \sim \epsilon,  
\quad \nabla \sim \epsilon, \quad \partial_t \sim \epsilon^2 .
\end{equation}
Balancing the linear term $\mu_2 f_2$ with the nonlinear term $f_1^* f_2$ also yields $\mu_2 \sim \epsilon^2$, which determines the relation between $\epsilon$ and the distance to the instability threshold.

\subsubsection{Polar particles with ferromagnetic interaction}

This case is slightly more complicated, as will appear below. Now the relevant order parameter is $f_1$.
We have to balance the following terms in the continuity equation (\ref{eq-continuity}) and in the equation for $f_1$ [see Eq.~(\ref{eq-fk})]
\begin{equation}
\partial_t \rho \sim {\rm Re}\nabla^* f_1, \qquad
\partial_t f_1 \sim \nabla \rho
\end{equation}
which implies $\rho-\rho_0 \sim f_1 \sim \epsilon$ and the propagative scaling
$\partial_t \sim \nabla \sim \epsilon^{\alpha}$.
If $\mu_2<0$, $f_2$ is slaved to $f_1$ and the linear term $\mu_2 f_2$
should be balanced by the nonlinear term $f_1^2$, yielding $f_2 \sim \epsilon^2$.
Finally, balancing the nonlinear term $f_1^* f_2$ with the diffusive term $\Delta f_1$ in the equation for $f_1$ leads to $\alpha =1$, resulting in the following scaling relations
\begin{equation} \label{eq-polar-scal}
f_1 \sim \epsilon, \quad f_2 \sim \epsilon^2, \quad \rho-\rho_0 \sim \epsilon, \quad  \nabla \sim \epsilon, \quad \partial_t \sim \epsilon.
\end{equation}
In addition, the relation $\mu_1 f_1 \sim f_1^* f_2$ yields $\mu_1 \sim \epsilon^2$.

This case is apparently quite similar to active nematics apart from the fact that the diffusive scaling $\partial_t \sim \nabla^2$ is replaced by the propagative one $\partial_t \sim \nabla$, but a major difference is that the resulting scaling does not allow all ``important terms'' to be balanced. This is true both in the continuity equation, where for instance $\Delta \rho \sim \epsilon^3$ while ${\rm Re}\nabla^* f_1 \sim \epsilon^2$, and most importantly in the equation for $f_1$, where terms like $\mu_1 f_1$ and $f_1^* f_2$ are of order $\epsilon^3$ while $\nabla \rho \sim \epsilon^2$. If one neglects the terms $\mu_1 f_1$ and $f_1^* f_2$, the resulting equation does not lead to collective order,
so that one has to keep terms of different order in the same equation to account for the relevant phenomenology.

Although this unbalance of terms may look surprising at first sight, the situation is similar for instance to what happens in the simple Fokker-Planck equation describing a biased random walk. The drift dynamics and the diffusive dynamics occur on different time scales, which implies that the drift and diffusive terms in the equation are generically of different order in an expansion in the lattice spacing. In the presence of a constant drift, one can reabsorb the drift term by going to a moving frame. In more complicated situations like the ones we are dealing with here, it is not easy to reabsorb the terms $\nabla \rho$ and ${\rm Re}\nabla^* f_1$ by a simple transformation. A proper way to deal with this issue would probably be to perform a multiscale expansion, introducing fast and slow time variables. Such a derivation however goes beyond the scope of the present paper.

\subsubsection{Polar particles with nematic interaction}

This last case shares similarities with both previous cases. The linearly unstable mode is $f_2$, which is thus the main order parameter of the problem, but one may also wish to keep $f_1$ as a relevant field because of the polarity of particles.
In this case, two different scalings can to some extent be considered as consistent.
The first one starts from the same reasoning as in the active nematics case (sect.~\ref{sec-active-nema-scal}), leading again to the scaling Eq.~(\ref{eq-active-nema-scal}).
Then, to determine the scaling of $f_1$ (which was absent from the
active nematics case), we balance the term $\partial_t \rho$
with ${\rm Re}\nabla^* f_1$, or equivalently $\partial_t f_2$
with $\nabla f_1$, leading to $f_1 \sim \epsilon^2$. Further, assuming
that $\mu_3$ and $\mu_4$ are negative so that $f_3$ and $f_4$ are
slaved to the nonlinear terms, one has $f_3 \sim f_1 f_2$ and $f_4
\sim f_2^2$, yielding $f_3 \sim \epsilon^3$ and $f_4 \sim \epsilon^2$.
Finally, balancing the lowest order nonlinear term coupling to higher
modes, $f_4 f_2^*$, with the diffusive term $\Delta f_2$ in the equation for $f_2$ leads to $\alpha =1$.
Altogether, we have
\begin{eqnarray} \nonumber
&& f_1 \sim \epsilon^2, \quad f_2 \sim \epsilon, \quad f_3 \sim \epsilon^3, \quad f_4 \sim \epsilon^2,\\
&& \quad \rho-\rho_0 \sim \epsilon, \quad \nabla \sim \epsilon, \quad \partial_t \sim \epsilon^2.
\label{eq-rods-scal-diff}
\end{eqnarray}
This scaling ansatz then leads to equations of the same form as the active nematics case. Under this scaling assumption, $f_1$ is purely slaved to $f_2$ (in the sense that $\partial_t f_1$ is a higher order term), and can eventually be eliminated from the equations.

An alternative view
is to consider that the propagative nature of the particles suggests a
ballistic scaling between time and space, namely $\partial_t \sim \nabla$. To fulfill this scaling in the continuity equation,
one then needs to assume that $f_1 \sim \epsilon$, making it a
relevant field. We thus end up with the following scaling ansatz,
\begin{eqnarray} \nonumber
&& f_1 \sim \epsilon, \quad f_2 \sim \epsilon, \quad f_3 \sim \epsilon^2, \quad f_4 \sim \epsilon^2,\\
&& \quad \rho-\rho_0 \sim \epsilon, \quad \nabla \sim \epsilon, \quad \partial_t \sim \epsilon
\label{eq-rods-scal}
\end{eqnarray}
leading to a richer set of closed equations, as shown in section \ref{rods}.

\subsubsection{Systematic general scaling ansatz}

The previous examples suggest the definition of a systematic scaling
ansatz valid in the generic case. First of all, density fluctuations
are of order $\epsilon$, as well as all Fourier modes up to the
lowest $k$ relevant order parameters, which we will denote as $k=h$ and
which is typically determined by the alignment symmetry, $h=2/n$ ($n=1$ or $2$).
The next $h$ modes then scale as $\epsilon^2$, and so on and so forth.
Space derivatives scaling is fixed by considering the equation for the relevant
order parameter(s). Balancing the diffusion term (of order
$\epsilon^{1+2\alpha}$) with the lowest order nonlinear terms coupling to higher
modes (always of order $\epsilon^{3}$), leads to $\alpha=1$ and $\nabla
\sim \epsilon$, consistently with the typical Ginzburg-Landau scaling
ansatz.
Finally, the scaling of time derivatives is fixed by the symmetry of
the displacement distribution $\Phi$, or equivalently, by the dominant
propagative mode in the continuity equation (\ref{eq-continuity}). Non
zero drift imposes the propagative scaling $\partial_t \sim \epsilon$,
while zero drift, diffusive dynamics leads to the diffusive scaling
$\partial_t \sim \epsilon^2$.

To summarize, we have for our systematic ansatz (with $h=2/n$)
\begin{eqnarray} 
\nonumber
&& \rho-\rho_0 \sim \epsilon, \quad f_{k\neq0} \sim \epsilon^{1+\lfloor (|k|-1)/h
  \rfloor}, \quad \nabla \sim \epsilon, \\
\label{general-scal}
&&\quad \quad \quad \quad \quad \quad \quad \quad {\rm and}\\
&& \partial_t \sim \epsilon\;{\rm (ballistic)} \quad
{\rm or} \quad \partial_t \sim \epsilon^2\; {\rm (diffusive)},
\nonumber
\end{eqnarray}
where $\lfloor \ldots \rfloor$ denotes the integer part.

\subsection{Truncation and closed hydrodynamic equations}
\label{truncation}

We now discuss the general structure of the closed equations emerging from a
truncation of Eqs.~(\ref{eq-fk}) by the general scaling ansatz
(\ref{general-scal}).

Independently of the truncation scheme, the dynamics of the density field fluctuations is given by the
continuity equation (\ref{eq-continuity}), which couples $\delta \rho$
to the polar (drift) and the nematic (anisotropic diffusion) order parameter
fields. The isotropic diffusion term $\sim \Delta \rho$, in the
absence of anisotropic diffusion, can be interpreted as of pure
thermal origin; otherwise, it carries a contribution from
nematic activity. Note that in `pure' self-propelled models, like the original VM, where displacement fluctuations
are ignored, diffusion contributions (both isotropic and anisotropic)
are absent.

Since additional symmetries in the system, and thus in the one-particle distribution $f$ manifest themselves by some zero Fourier
coefficients, one can check from Eq.~(\ref{eq-continuity}) that global symmetries of higher than nematic order results in both $f_1=0$ and
$f_2=0$, completely decoupling density from order parameter dynamics.
This suggests that, at least in two spatial dimensions, no
out-of-equilibrium dynamics may emerge from active particles with
global symmetries higher than the nematic one, so that $m=1,2$ are the
only relevant symmetries.

We now turn our attention to the order parameter equations of order
$k\geq1$. They couple the mode $f_k$ to terms of order $k \pm 1,2$
via linear spatial derivatives, and to terms of all orders via the
nonlinear coupling term
\begin{equation}
\mathcal{C}_k=\sum_{q=-\infty}^{\infty} ({P}_k I_{k,q}-I_{0,q}) f_q f_{k-q}\,.
\end{equation}
Terms of order $q=0,k$ in the sum can be singled out to yield the
density dependent linear coefficient 
\begin{equation} \label{def-muk2}
\mu_k = -(1-{P}_k) +\omega_k \rho
\end{equation}
which controls the local stability of the disordered phase\footnote{Note that, crucially, density dependence cancels out for metric-free
systems.}. 
In the following we assume that all linear coefficients $\mu_k$ are
negative except mode $h=2/n$, associated to the 
alignment symmetry and relevant order parameter, which for high density and/or
low noise may turn positive, $\mu_h>0$.

We have seen above that the first non-linear terms
appearing in the relevant order parameter equation are typically of order
$\epsilon^3$. As a result, we expand all equations to order
$\epsilon^3$. This results in a well-controled truncation close to the
instability threshold, in such a way that equations for additional
modes which are not retained as order parameters
provide closure relations for the order parameter equations.

The relevant order parameter equation couples $f_h$ to higher
order terms via order $\epsilon^3$ terms of $\mathcal{C}_h$ of the kind
$f_{h+j} f_j^*$, with $j=1,\ldots,h$. Higher order terms, in turn, are
slaved to order $\epsilon$ modes since $\mu_k<0$ and 
$|\partial_t f_{k}| \ll |\mu_{k} f_{k}|$. 
In particular, for $k=2h$ (the first higher non-zero mode when $m=n$)
one has to order $\epsilon^2$ (cubic terms may be ignored since inserting them in the nonlinear
coupling term in the equation for $f_h$ would result in terms $\sim \epsilon^4$)
\begin{equation} 
f_{2h} \approx -\frac{1}{\mu_{2h}}\left( {P}_{2h} I_{2h,h}-I_{0,h}\right) f_h^2
-\delta_{h,1} \frac{v_0}{2} \nabla f_1
\label{slaved}
\end{equation}
with the gradient term appearing only for the polar case $h=1$.
Substituting the slaved mode $f_{2h}$ into the $k=h$ equation
one obtains the saturating cubic term $-\xi_h f_h
|f_h|^2$ with the coefficient
\begin{equation}
\xi_h = \frac{{P}_{2h} I_{2h,h}-I_{0,h}}{\mu_{2h}}\left[{P}_h (I_{h, 2h} + I_{h, -h}) - I_{0, 2h} -I_{0,-h}\right]
\end{equation}
being typically positive in the parameter region where $\mu_h>0$.

If $m=n$ the hydrodynamic equations have a relatively simple
structure, consisting of the continuity equation
(\ref{eq-continuity}) and the relevant order parameter equation of the
form
\begin{equation}
\frac{\partial f_h}{\partial t} = \left[\mu_h(\sigma, \rho) -\xi_h
  |f_h|^2\right]f_h + \nabla (\dots)+ \nabla\nabla (\dots)
\label{eq-op}
\end{equation}
where $\nabla$ and $\nabla\nabla$ indicate generic spatial derivatives
(respectively convective and 
diffusive) of the density $\rho$ and order parameter
$f_h$ fields. $\nabla$ terms appear only for $m=1$ (convective terms
are forbidden in a nematic $m=2$ or higher symmetry) and, due to the substitution of the slaved higher order term
(\ref{slaved}), are both linear and nonlinear.  On the other hand, $\nabla\nabla$ terms
are linear to order $\epsilon^3$. Obviously, all terms in Eq.~(\ref{eq-op}) need to respect the rotational invariance of the system.

The familiar Ginzburg-Landau term $\left[\mu_h(\sigma, \rho) -\xi_h
  |f_h|^2\right]f_h$ appearing in Eq.~(\ref{eq-op}) locally describes the spontaneous breaking of a
continuous symmetry, with the ordered solution
\begin{equation}
|f_h|=\sqrt{\frac{\mu_h}{\xi_h}}.
\label{w1}
\end{equation}

On the other hand, if $n>m$, the order parameter fields $k<h$ are not
zero by symmetry, and by our scaling ansatz (\ref{general-scal}) their dynamics may be
relevant. Their equations need to be
added to the hydrodynamic description. Moreover, the modes $f_j$ with
$j=1,\ldots, h$ are now generically coupled to the enslaved modes 
$f_{h+j}$. Substituting the latter in the former, generates a larger
number of nonlinear terms, which now also include different
combinations of the fields without spatial derivatives.
However, the local symmetry breaking phenomenon is still described by
the Ginzburg-Landau term appearing in the $f_h$ equation.

For completeness, in the following, we derive the simplest well-behaved
equations describing the physics at hand for the three basic classes
introduced above.

\subsubsection{Nematic particles with nematic interactions}

In the case of nematic particles, equations simplify since $v_0=0$ and $f_1=0$.
The scaling of the different fields is given by Eq.~(\ref{eq-active-nema-scal}).
First of all, the continuity equation (\ref{eq-continuity}) reads in this case, given all non-zero terms are of order $\epsilon^3$,
\begin{equation} \label{eq-continuity-nema}
\frac{\partial \rho}{\partial t}  = 
D_0 \Delta \rho
+ \frac{D_1}{2} {\rm Re} (\nabla^{*2} f_2).
\end{equation}
The equation for $f_2$ reads, to order $\epsilon^3$,
\begin{equation}
\frac{\partial f_2}{\partial t} = \mu_2 f_2 + D_0 \Delta f_2 + \frac{D_1}{4} \nabla^2 \rho
+ [{P}_2 (I_{2,4}+I_{2,-2})-I_{0,4}-I_{0,-2}] f_2^* f_4 .
\label{eq-f2-nema0}
\end{equation}
Similarly, the equation for $f_4$ up to order $\epsilon^3$ is given by
\begin{equation}
\frac{\partial f_4}{\partial t} = \mu_4 f_4 + \frac{D_1}{4} \nabla^2 f_2 + ({P}_4 I_{4,2}-I_{0,2}) f_2^2 .
\end{equation}
We recall that the coefficients $\mu_k$ are defined in Eq.~(\ref{def-muk2}).
Assuming $\mu_4<0$, which has to be checked case by case (it is indeed true for the representative member of the class with $H(\Delta)=\frac{\Delta}{2}$ \cite{ACTIVE-NEMA}), $f_4$ can be slaved to $f_2$. This slaving procedure is indeed consistent with scaling considerations. Since $f_4$ appears in Eq.~(\ref{eq-f2-nema0}) only in the product $f_2^* f_4$, only terms up to order $\epsilon^2$ need to be kept in the expression of $f_4$, leading to
\begin{equation}
f_4 = -\frac{1}{\mu_4} ({P}_4 I_{4,2}-I_{0,2}) f_2^2.
\end{equation}
Injecting this expression of $f_4$ into Eq.~(\ref{eq-f2-nema0}), one finds the following closed equation for $f_2$,
\begin{equation}
\frac{\partial f_2}{\partial t} = \mu_2 f_2 + \nu \Delta f_2 + \chi \nabla^2 \rho - \xi |f_2|^2 f_2,
\end{equation}
where the coefficients $\nu$, $\chi$ and $\xi$ are defined as
\begin{eqnarray}
\nu &=& D_0,\quad \chi = \frac{D_1}{4} \\
\xi &=& \frac{{P}_4 I_{4,2}\!-\!I_{0,2}}{\mu_4}[{P}_2 (I_{2,4}\!+\!I_{2,-2})\!-\!I_{0,4}\!-\!I_{0,-2}] . \nonumber
\end{eqnarray}
Note that here, all terms allowed by symmetry\footnote{It is useful to specify how the "symmetry" of terms is characterized. In the same way as for vectors and tensors, it describes how a given quantity transforms under a rotation of the frame axis. Under a rotation of angle $\theta_0$ of the frame axis, the distribution $f(\theta)$ is changed into $f(\theta-\theta_0)$, so that in turn the Fourier coefficients $f_k$ are changed into $e^{ik\theta_0} f_k$. More generally, if a quantity is multiplied by a factor $e^{is\theta_0}$ under a rotation of axis of angle $\theta_0$, we shall call 'spin' the integer $s$. Hence $f_k$ has a spin $s=k$ (note that $s$ can be negative). As expected, the polar order parameter $f_1$ has spin $1$ and maps to a vector, while the nematic order parameter has spin $2$ and maps to a tensor.
The derivation operator $\nabla$ has a spin $s=1$, and complex conjugation reverses the sign of $s$. 
The total spin of a product of factors is the sum of the spin of each factor.
In the evolution equation for $f_k$, the only terms allowed by symmetry are thus those with a total spin equal to $k$.}
up to order $\epsilon^3$ have been generated.

\subsubsection{Polar particles with ferromagnetic interactions}

For polar particles, the scaling of the different quantities is given in Eq.~(\ref{eq-polar-scal}).
Retaining terms up to order $\epsilon^3$, the continuity equation reads
\begin{equation} \label{eq-continuity-polar}
\frac{\partial \rho}{\partial t} + v_0 {\rm Re} (\nabla^* f_1) = 
D_0 \Delta \rho.
\end{equation}
Truncating the equation for $f_1$ to order $\epsilon^3$ yields
\begin{eqnarray} \label{eq-f1-polar0}
\frac{\partial f_1}{\partial t} + \frac{v_0}{2} \left( \nabla \rho + \nabla^* f_2\right) &=& \mu_1 f_1 + D_0 \Delta f_1 + \frac{D_1}{4} \nabla^2 f_1^* \\ \nonumber
&+& [{P_1} (I_{1,-1}+I_{1,2}) - I_{0,-1} - I_{0,2}] f_1^* f_2 .
\end{eqnarray}
Since $f_2$ appears only in the terms $\nabla^* f_2$ and $f_1^* f_2$, only terms of order up to $\epsilon^2$ need to be retained in the expansion of $f_2$, yielding
\begin{equation} \label{eq-f2-polar}
\frac{v_0}{2} \nabla f_1 = \mu_2 f_2 + ({P}_2 I_{2,1} - I_{0,1}) f_1 ^2 .
\end{equation}
Note that the time derivative has been dropped for being of order $\epsilon^3$. As already outlined in the active nematics case, this procedure is however only consistent when the linear coefficient $\mu_2$ is negative, so that $f_2$ can be slaved to $f_1$.

From Eq.~(\ref{eq-f2-polar}), $f_2$ can be expressed as a function of $f_1$ and $\nabla f_1$. Injecting the resulting expression for $f_2$
in Eq.~(\ref{eq-f1-polar0}), one eventually finds
\begin{equation}
\frac{\partial f_1}{\partial t} = -\frac{v_0}{2}\nabla \rho + \mu_1 f_1 - \xi |f_1|^2f_1 + \nu \Delta f_1
+ \chi \nabla^2 f_1^* + \kappa_1 f_1\nabla^* f_1 + \kappa_2 f_1^* \nabla f_1
\end{equation}
where the coefficients are given by
\begin{eqnarray}
\nu &=& D_0 +\frac{v_0^2}{4\,|\mu_2|},\quad
\chi = \frac{D_1}{4} \nonumber\\
\kappa_1 &=& \frac{v_0}{\mu_2}({P}_2 I_{2,1}-I_{0,1}) \\
\kappa_2 &=& \frac{v_0}{2\mu_2} [{P}_1 (I_{1,-1}+I_{1,2}) - I_{0,-1} - I_{0,2}]\nonumber\\
\xi &=& \frac{{P}_2 I_{2,1}\!-\! I_{0,1}}{\mu_2}
[{P}_1 (I_{1,-1}\!+\!I_{1,2})\!-\!I_{0,-1}\!-\!I_{0,2}]\nonumber
\end{eqnarray}
the coefficient $\mu_2$ being given by Eq.~(\ref{def-muk2}) with $k=2$.

Note that contrary to the case of particles with continuous time ballistic motion along their polarity vector, as studied in \cite{BDG,TOPOKINETIC}, the presence of diffusion at the microscopic level of the dynamics generates an anisotropic diffusion term $\nabla^2 f_1^*$. Only the term $f_1\nabla f_1^*$, that would be allowed by symmetry, is not present here, and can thus not be obtained from point-like particles with (purely local) binary interactions.

\subsubsection{Polar particles with nematic interactions}
\label{rods}

For polar particles with nematic interactions (``rods''), the scaling of the different quantities is given in Eq.~(\ref{eq-rods-scal}). From this scaling, one sees that all terms in the continuity equation (\ref{eq-continuity}) have to be kept.
After truncation to order $\epsilon^3$, the equation for $f_1$ reads, using Eq.~(\ref{eq-fk}),
\begin{eqnarray} \nonumber
\frac{\partial f_1}{\partial t} &+& \frac{v_0}{2}(\nabla \rho + \nabla^* f_2) = \mu_1 f_1 + D_0 \Delta f_1 + \chi \nabla^2 f_1^*\\
&+& [{P_1} (I_{1,-1}+I_{1,2}) - I_{0,-1} - I_{0,2}] f_1^* f_2
+ [{P_1} (I_{1,-2}+I_{1,3}) - I_{0,-2} - I_{0,3}] f_2^* f_3
\label{eq-f1-rods0}
\end{eqnarray}
with $\chi = \frac{D_1}{4} $ as above.
Similarly, one finds for $f_2$
\begin{eqnarray} \nonumber
\frac{\partial f_2}{\partial t} &+& \frac{v_0}{2}(\nabla f_1 + \nabla^* f_3) = \mu_2 f_2 + D_0 \Delta f_2 + \chi \nabla^2 \rho
+ ({P_2} I_{2,1} - I_{0,1}) f_1^2 \\
&+& [{P_2} (I_{2,-1}+I_{2,3}) - I_{0,-1} - I_{0,3}] f_1^* f_3
+ [{P_2} (I_{2,-2}+I_{2,4}) - I_{0,-2} - I_{0,4}] f_2^* f_4 \, .
\label{eq-f2-rods0}
\end{eqnarray}
One also needs to consider the equations for $f_3$ and $f_4$. Since $f_3$ and $f_4$ appear combined with $f_1$, $f_2$ or derivative operators, we only need to expand these two modes to order $\epsilon^2$, leading to
\begin{eqnarray}
\label{eq-f3-rods}
f_3 &=& \frac{v_0}{2\mu_3} \nabla f_2 \!-\! \frac{1}{\mu_3} [{P_3} (I_{3,1}\!+\!I_{3,2}) \!-\! I_{0,1} \!-\! I_{0,2}] f_1 f_2\\
f_4 &=& -\frac{1}{\mu_4} ({P}_4 I_{4,2} - I_{0,2}) f_2^2 \, .
\label{eq-f4-rods}
\end{eqnarray}
Note that, here again, enslaving $f_3$ and $f_4$ to $f_1$ and $f_2$ is only consistent
if $\mu_3$ and $\mu_4$ are negative.
Injecting these expressions of $f_3$ and $f_4$ in Eqs.~(\ref{eq-f1-rods0}) and (\ref{eq-f2-rods0}) for $f_1$ and $f_2$, we end up with
\begin{eqnarray}
\label{eq-f1-rods}
\frac{\partial f_1}{\partial t} &=& - \frac{v_0}{2} (\nabla \rho + \nabla^* f_2)
+ \mu_1 f_1 + \beta |f_2|^2 f_1 + \zeta f_1^* f_2 \\ \nonumber
&& \qquad \qquad +\nu_1 \Delta f_1 + \chi \nabla^2 f_1^* + \gamma f_2^*\nabla f_2\\ \nonumber
\frac{\partial f_2}{\partial t} &=& - \frac{v_0}{2} \nabla f_1
+ (\mu_2 -\xi |f_2|^2) f_2 + \omega f_1^2 + \tau |f_1|^2 f_2\\
&& \qquad \qquad +\nu_2 \Delta f_2 + \chi \nabla^2 \rho  + \kappa_1 f_1^* \nabla f_2 + \kappa_2 \nabla^* (f_1 f_2)
\label{eq-f2-rods}
\end{eqnarray}
where the coefficients are given by
\begin{eqnarray}
\gamma &=& \frac{v_0}{2\mu_3} [{P}_1 (I_{1,-2}+I_{1,3}) - I_{0,-2} - I_{0,3}]\nonumber\\
\beta &=& -\frac{2\gamma}{v_0} [{P}_3 (I_{3,1}+I_{3,2}) - I_{0,1} - I_{0,2}]\nonumber\\
\zeta &=& {P}_1 (I_{1,-1}+I_{1,2}) - I_{0,-1} - I_{0,2}\nonumber\\
\nu_1 &=& D_0, \qquad
\chi = \frac{D_1}{4} \nonumber\\
\nu_2 &=& D_0 + \frac{v_0^2}{4\,|\mu_3|}, \quad \omega = {P}_2 I_{2,1} - I_{0,1}\\
\kappa_1 &=& \frac{v_0}{2\mu_3} [{P}_2 (I_{2,-1}+I_{2,3}) - I_{0,-1} - I_{0,3}]\nonumber\\
\kappa_2 &=& \frac{v_0}{2\mu_3} [{P}_3 (I_{3,1}+I_{3,2}) - I_{0,1} - I_{0,2}]\nonumber\\
\xi &=& \frac{{P}_4 I_{4,2} \!-\! I_{0,2}}{\mu_4} [{P}_2 (I_{2,-2}\!+\!I_{2,4}) \!-\! I_{0,-2} \!-\! I_{0,4}]\nonumber\\
\tau &=& -\frac{2\kappa_1}{v_0} [{P}_3 (I_{3,1}+I_{3,2}) - I_{0,1} - I_{0,2}] \, . \nonumber
\end{eqnarray}
Although Eqs.~(\ref{eq-f1-rods}) and (\ref{eq-f2-rods}) already contain many terms, not all terms allowed by symmetry are present, especially in the equation for $f_1$. Note however that if one further enslaves $f_1$ to $f_2$, new terms will be generated in the equation for $f_2$.


\section{Discussion}
\label{S4}

\subsection{$\rho$-dependence}

In Section \ref{S3}, we derived a number of transport coefficients appearing in the hydrodynamic equations for the three main classes of models.
Although this was not made explicit, most of the coefficients we have derived actually depend on the {\it local} density $\rho$, through the linear coefficients $\mu_k$. A natural question is then that of the qualitative influence of this
dependence of the transport coefficients on the density field. Indeed, it would be more convenient to have constant coefficients, depending only on microscopic parameters of the model and on the mean density. As we will see below, the $\rho$ dependence of 
the linear term governing the instability of the order parameter is at the origin of the linear instability of the homogeneous 
ordered state ultimately leading to the emergence of the inhomogeneous solutions in the transition region. 
It is thus essential and must be kept.
The density-dependence of the other linear terms $\mu_k f_k$, in contrast, is only felt 
indirectly in the definition of the coefficients of the other terms.  It is tempting to expand them around the mean density $\rho_0$, 
setting $\rho=\rho_0+\delta\rho$. For terms of order $\epsilon^3$, corrections in $\delta\rho$ of the 
coefficients lead to terms of order $\epsilon^4$ or higher, and can thus a priori be discarded.

The question however becomes more subtle if one wishes to assess the qualitative validity of the derived hydrodynamic equations further away from the threshold. Strictly speaking, these equations are of course not expected to be valid away from the threshold, which could in a sense close the debate. Yet, the example of Ginzburg-Landau equations shows that in many cases the qualitative behavior away from the linear instability threshold remains well described by the obtained equations.
Unfortunately, there is no well-defined general procedure to discriminate between different formulations of the same terms, if these formulations become equivalent close to threshold. This is the case in particular for the density-expansion of the transport coefficients, and one has to rely on numerical integrations of the hydrodynamic equations to check the effect of the $\rho$ dependence of the different terms.

For the coefficient $\xi$ of the cubic term, some theoretical justifications of the necessity to keep the full $\rho$-dependence of $\xi$ (as opposed to a dependence on the mean density $\rho_0$) close to threshold can also be provided. 
For instance, the role of the density-dependence of $\xi$ appears in the dynamical system analysis of the shape of non-linear patterns arising for polar particles interacting ferromagnetically \cite{BARTOLO}.
Note also that the dependence of $\xi$ over the average density $\rho_0$, that we keep when doing the expansion of the density field around the mean density value, is crucial for the value $f_k^{(0)}/\rho=\rho^{-1} \sqrt{\mu_k/\xi}$ of the polarity (in the homogeneous phase) to increase with mean density and to saturate to a finite value at large mean density.
Finally, when studying the linear instability of the homogeneous ordered state, 
one finds that the term in $d\xi/d\rho$ (obtained from a density expansion of $\xi(\rho)$) is not negligible (with respect to other terms that have to be kept) in the computation of the growth rates, and may lead to significant difference in the linear stability diagrams. We illustrate this in the next section.


\begin{figure}[t!]
\begin{center}
\includegraphics[width=\columnwidth]{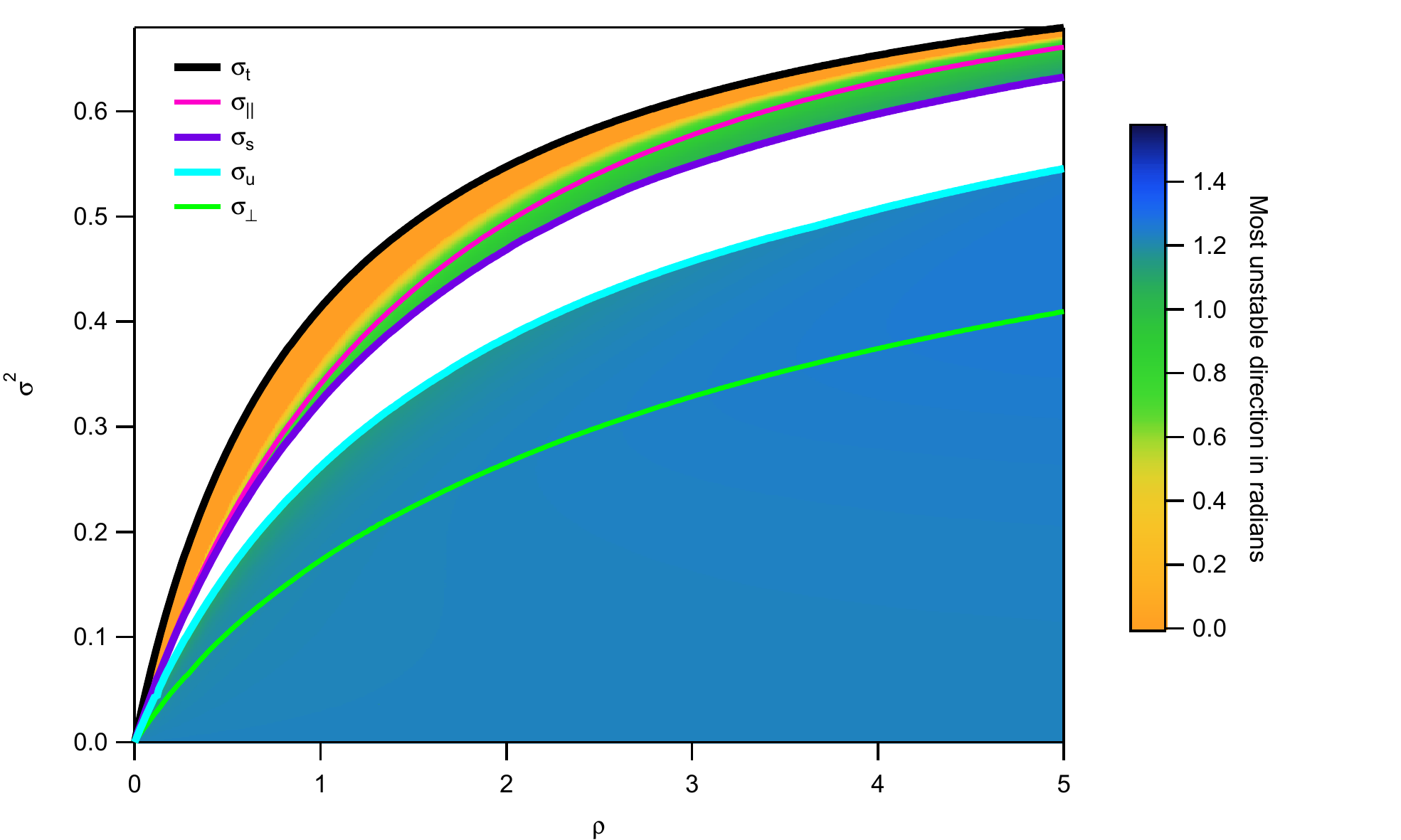}
\caption{
Phase diagram of the homogeneous solutions of the hydrodynamic equations for polar particles with ferromagnetic interactions in the noise-density plane, in the absence of microscopic positional diffusion ($D_0=D_1=0$), and without the $\rho$ dependence of the transport coefficients of nonlinear terms.
The full black line $\sigma_t$ indicates the stability limit of the homogeneous disordered state (upper white area). The lower white area corresponds to the stability domain of the homogeneous ordered state, delimited from above by the line $\sigma_s$ and from below by the line $\sigma_u$. In colored regions, the homogeneous ordered state is unstable. The color codes for the angle between the most unstable wavevector and the direction of order.
The line $\sigma_{||}$ indicates the limit above which longitudinal (i.e., with wavevector parallel to order) perturbations are unstable (the upper limit being $\sigma_t$). The line $\sigma_{\perp}$ indicates the limit below which transverse (i.e., with wavevector perpendicular to order) perturbations are unstable.
}
\label{diag-stab-polar}
\end{center}
\end{figure}



\begin{figure}[t!]
\includegraphics[width=0.5\columnwidth]{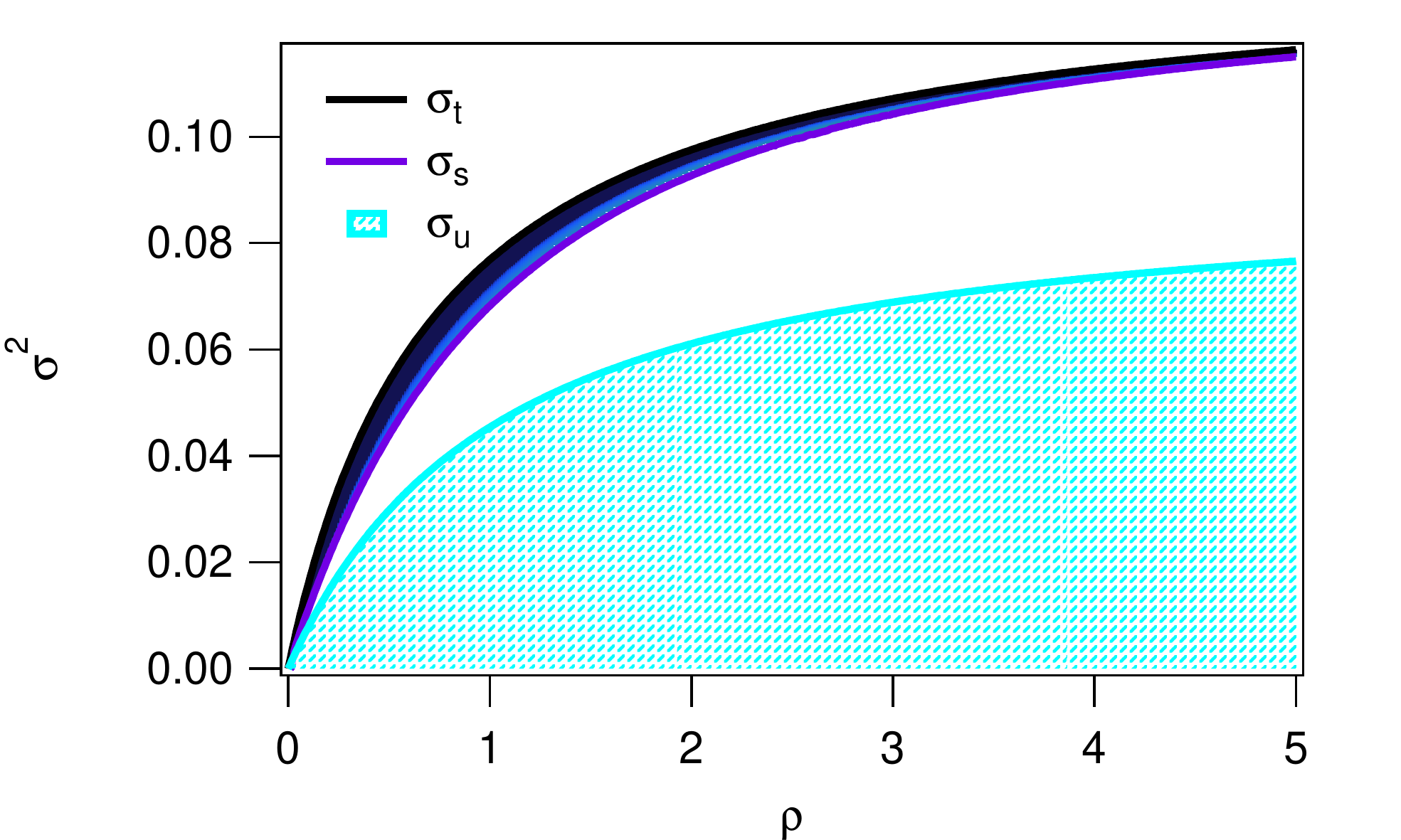}
\hfill
\includegraphics[width=0.5\columnwidth]{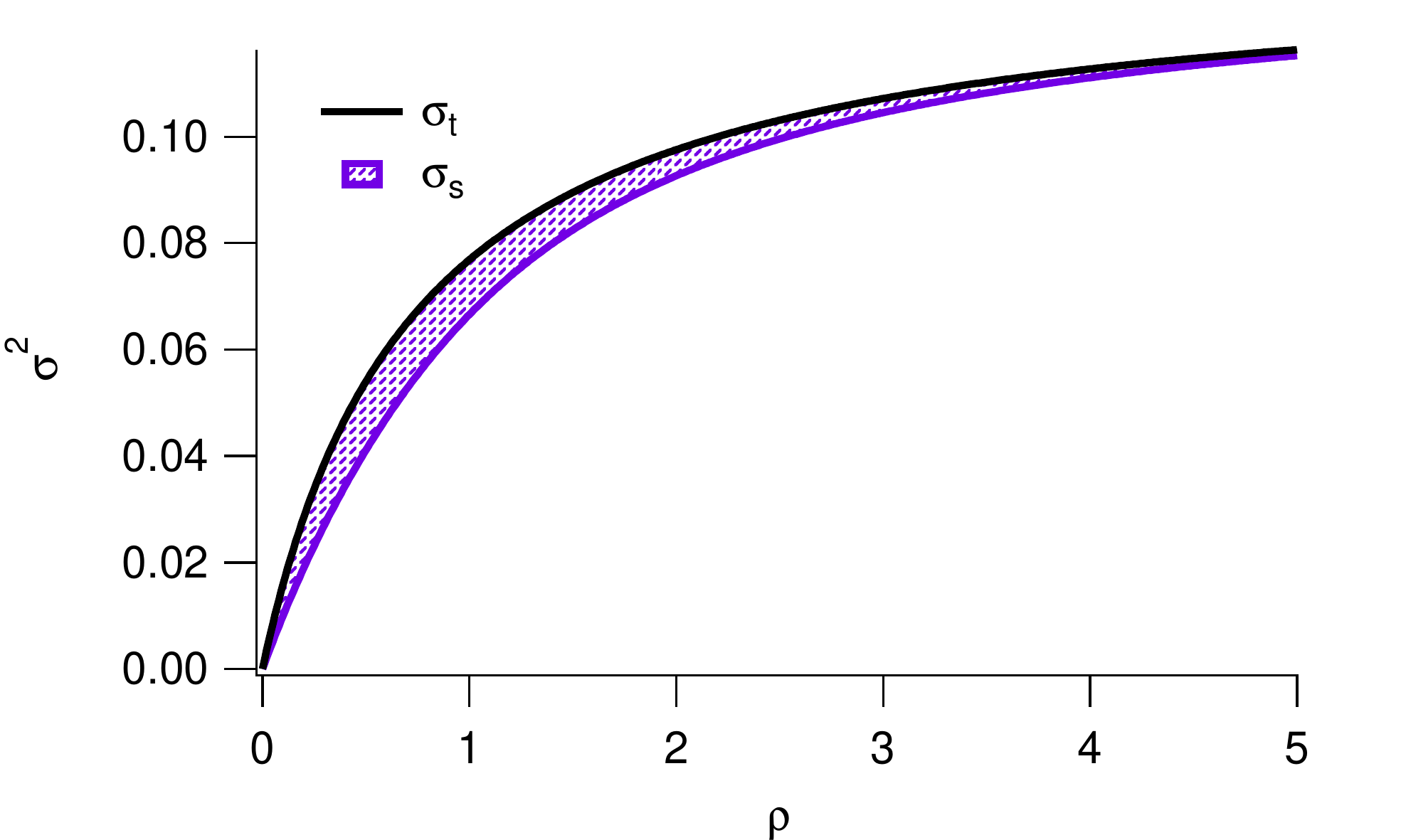}\\
\includegraphics[width=0.5\columnwidth]{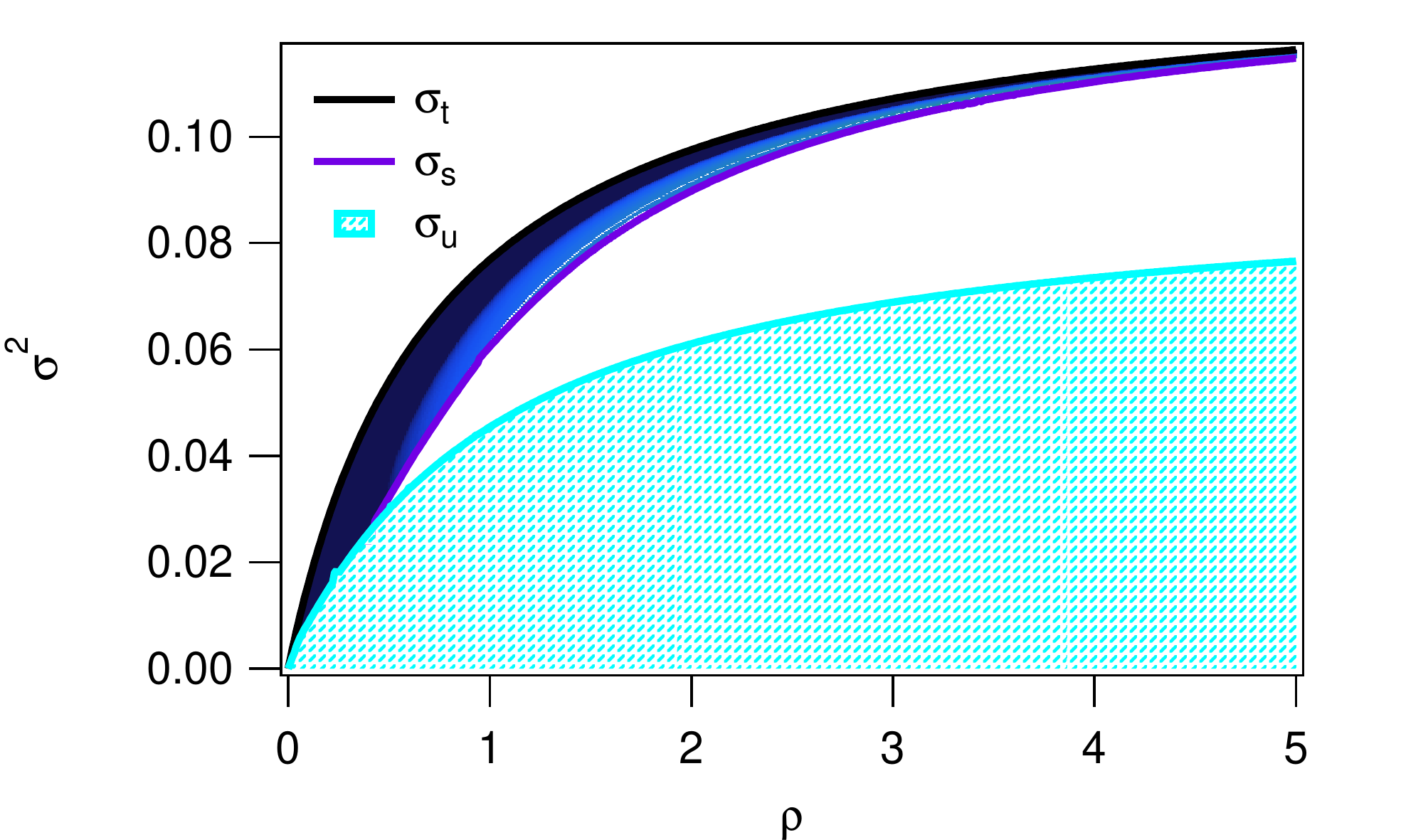}
\hfill
\includegraphics[width=0.5\columnwidth]{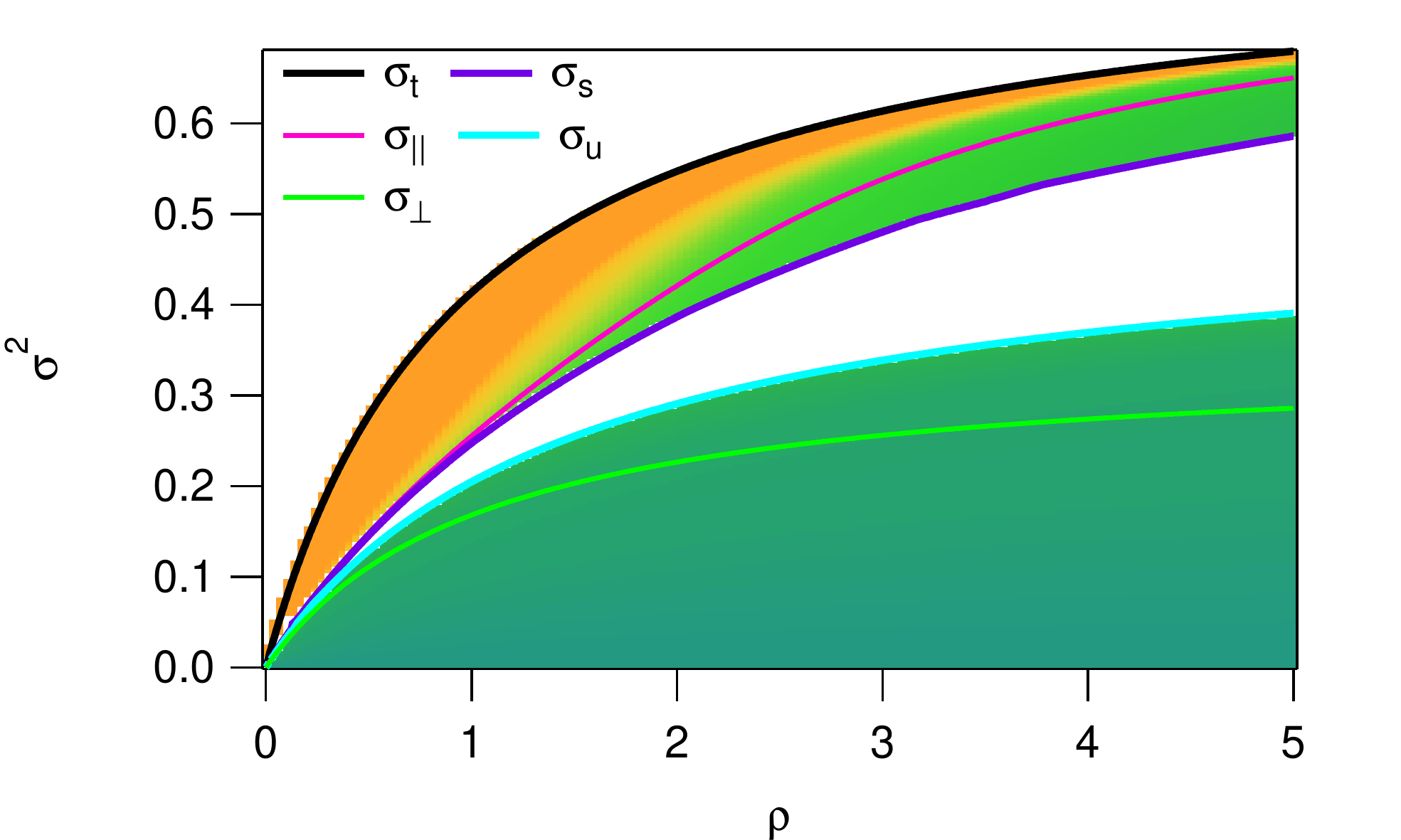}
\caption{
Phase diagram of the linear stability analysis of the homogeneous solutions to the hydrodynamic equations with and without
the $\rho$-dependence of the transport coefficients of the nonlinear terms (see text).
Left column: polar particles aligning nematically, without and with $\rho$ dependence (top and bottom panel respectively).
Top right: nematic particles aligning nematically (no discernible influence of $\rho$ dependence).
Bottom right: polar particle aligning ferromagnetically with $\rho$ dependence (compare with Fig.~\ref{diag-stab-polar}).
No positional diffusion ($D_0=D_1=0$). Same colormap, same legends as in Fig.~\ref{diag-stab-polar}.
}
\label{diag-stab-others}
\end{figure}


\subsection{Linear stability of homogeneous solutions}

Denoting as $f_{h}$ the relevant order parameter, either polar ($h=1$) or nematic ($h=2$), an ordered solution generically exists in the region where the disordered state becomes unstable ($\mu_{h}>0$).
We denote this line $\sigma_t(\rho)$ in the basic noise-density phase diagram in which we performed a systematic study of the linear instability of the homogeneous ordered solution, not restricting ourselves to modes longitudinal or transversal to order.
Results below illustrate this analysis. 

In previous studies of the cases without positional diffusion, which in the case of polar particles yield slightly simpler equations (anisotropic diffusion terms like $\nabla^2 f_1^*$ are absent due to a continuous time microscopic dynamics), a generic qualitative scenario was found in all cases, irrespective of the symmetries of the particles and of the interactions \cite{BDG,RODS-KINETIC,ACTIVE-NEMA}. 
We first illustrate it in the case of polar particles aligning ferromagnetically where only the $\rho$ dependence of the linear term has been kept (Fig.~\ref{diag-stab-polar}).

\subsubsection{General scenario}
\label{sect-general-scenario}

Close to the line $\sigma_t$ corresponding to the onset of order, the homogeneous ordered state is unstable with respect to long-wavelength perturbations, leading to the formation of non-linear patterns (see next section). This instability is generically related to the density-dependence of the linear coefficient $\mu_h$, the latter generically being an increasing function of the local density $\rho$ (only in the metric-free case, where $\mu_h$ becomes independent of the density, does one recover a stable homogeneous phase close to the transition line \cite{TOPOKINETIC}).

This instability region bordering the line $\sigma_t(\rho)$ closes at the line $\sigma_s(\rho)$
below which the homogeneous ordered solution is linearly stable (Fig.~\ref{diag-stab-polar}).
However, there exists a second instability region below a line $\sigma_u(\rho)$.
Preliminary studies of the stability of the Boltzmann equation in the case of ferromagnetic interactions indicate that this second instability region does not exist at the Boltzmann level. We thus conclude that this second instability region is an artifact of the truncation procedure.

For the unstable region between $\sigma_t$ and $\sigma_s$, the unstable modes have long wavelengths and are oriented 
mostly along order, but not perfectly so. The line marking the limit of stability of purely longitudinal modes is marked in magenta.
The second, "spurious" instability region below  $\sigma_u$ has large growth rates and unstable wavevectors mostly transversal to order, although again not perfectly so (the line marking the onset of purely transversal unstable modes is marked in green inside this region).

For the case of particles aligning nematically, the unstable modes are almost transversal in the first region.
The second region does not exist for active nematics, while its most unstable mode is at $k=0$ for the self-propelled rods case
 (Fig.~\ref{diag-stab-others}, top panels).

\subsubsection{Influence of the $\rho$-dependence of transport coefficients}

We now illustrate the influence of keeping the local density dependence in the transport coefficients, and notably in the cubic nonlinear term. 
For polar particles aligning ferromagnetically, this widens the region of stability of the homogeneous ordered state, 
and reduces the extent of the second instability region (compare Fig.~\ref{diag-stab-polar} and Fig.~\ref{diag-stab-others}, bottom right).

For the self-propelled rods case, keeping the $\rho$ dependence has a dramatic effect: the second region now "invades" the first one near the origin, becoming dominant (compare top left and bottom left panels in Fig.~\ref{diag-stab-others}). 

For active nematics, the $\rho$ dependence has a very weak effect, not discernable on the scale of Fig.~\ref{diag-stab-others} (see top right panel).

\subsubsection{Influence of positional diffusion}

We finally illustrate the influence of positional diffusion (coefficients $D_0$ and $D_1$) on the linear stability diagram.
As expected, the general effect of diffusion is to stabilize the homogeneous ordered solution. The extent of the first instability region
(between lines $\sigma_t$ and $\sigma_s$) varies only weakly,  whereas the second, spurious, region can be made very small 
under the influence of strong diffusion. We now illustrate these findings for the case of polar particles aligning ferromagnetically. 

\begin{figure}[t!]
\includegraphics[width=0.5\columnwidth]{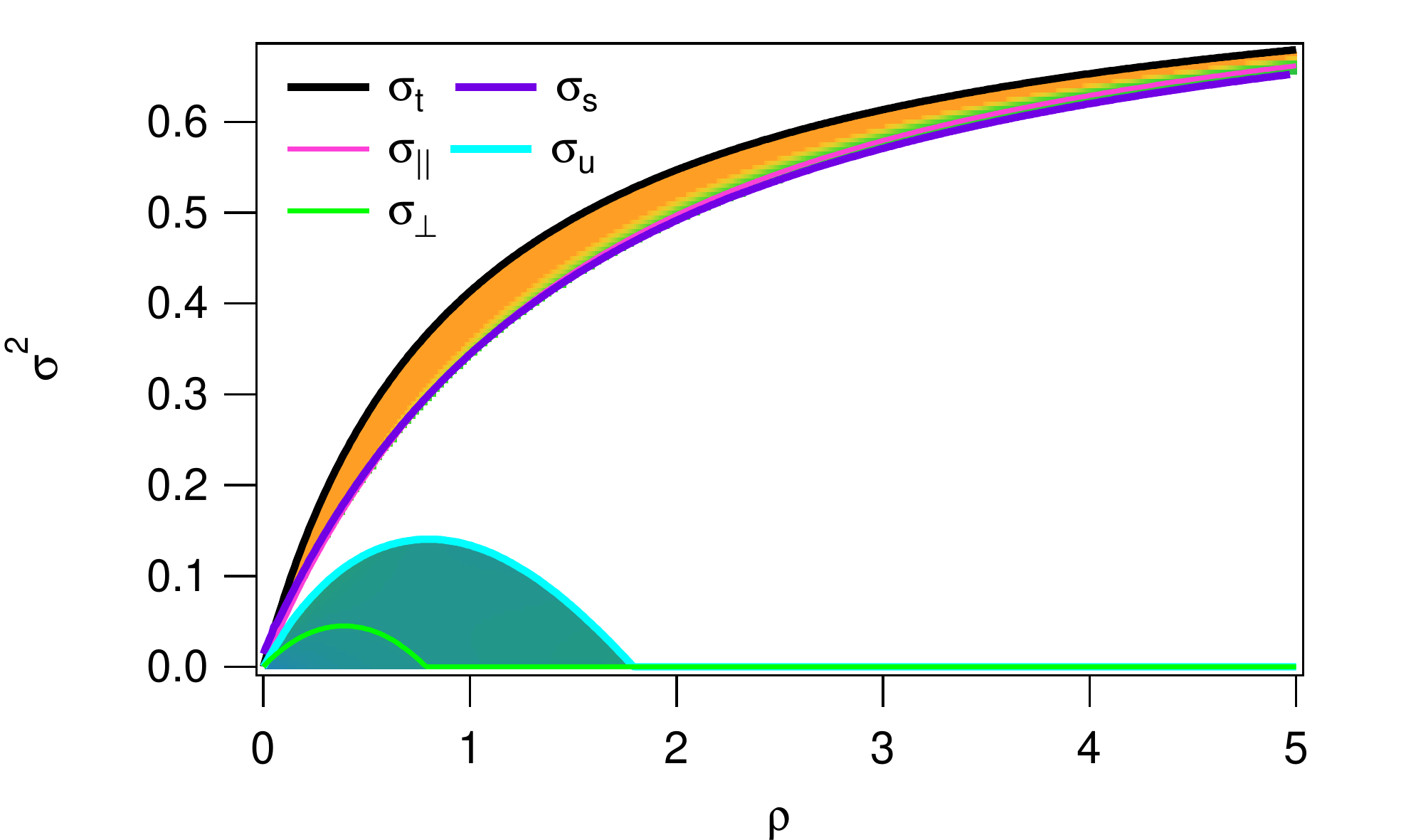}
\hfill
\includegraphics[width=0.5\columnwidth]{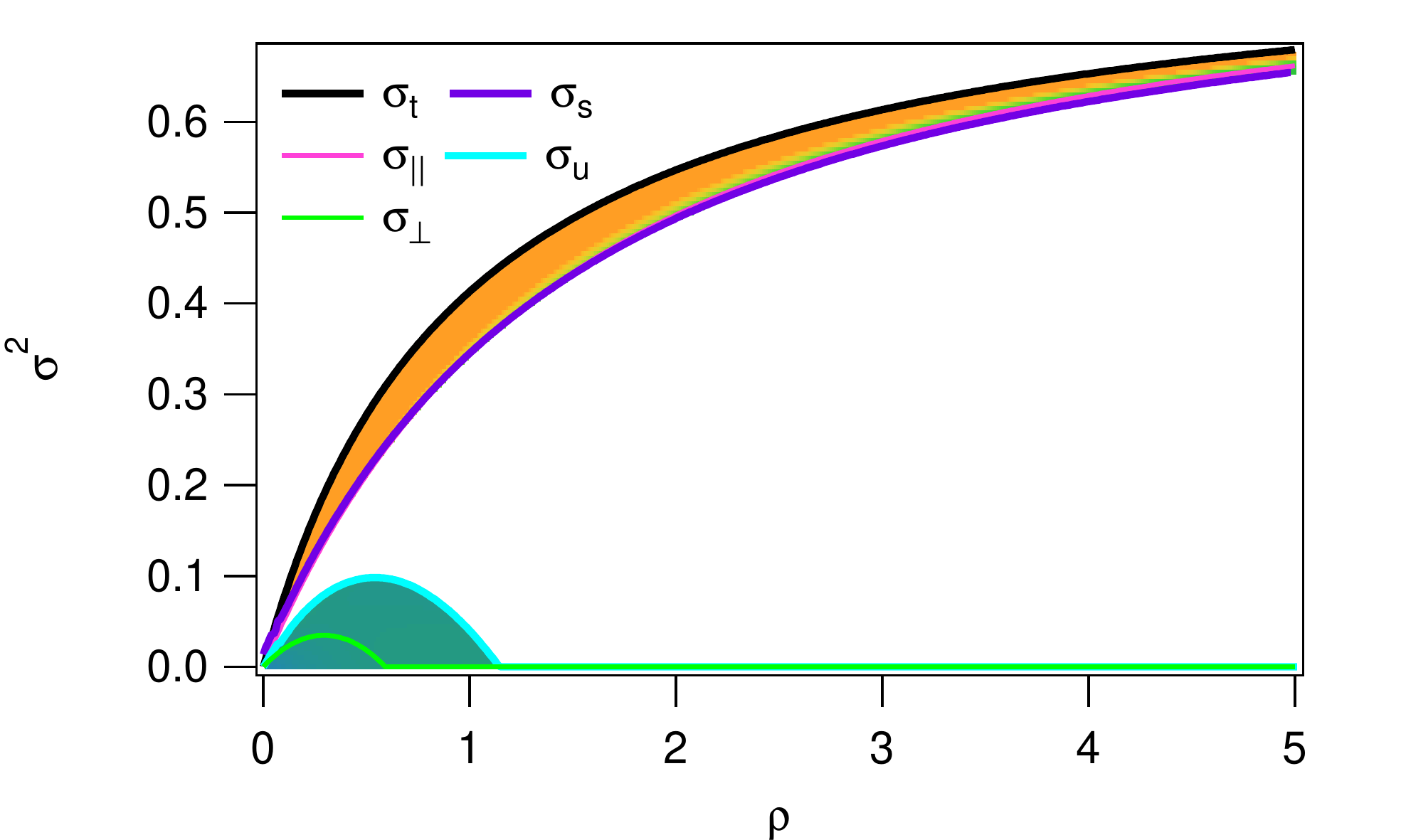}\\
\includegraphics[width=0.5\columnwidth]{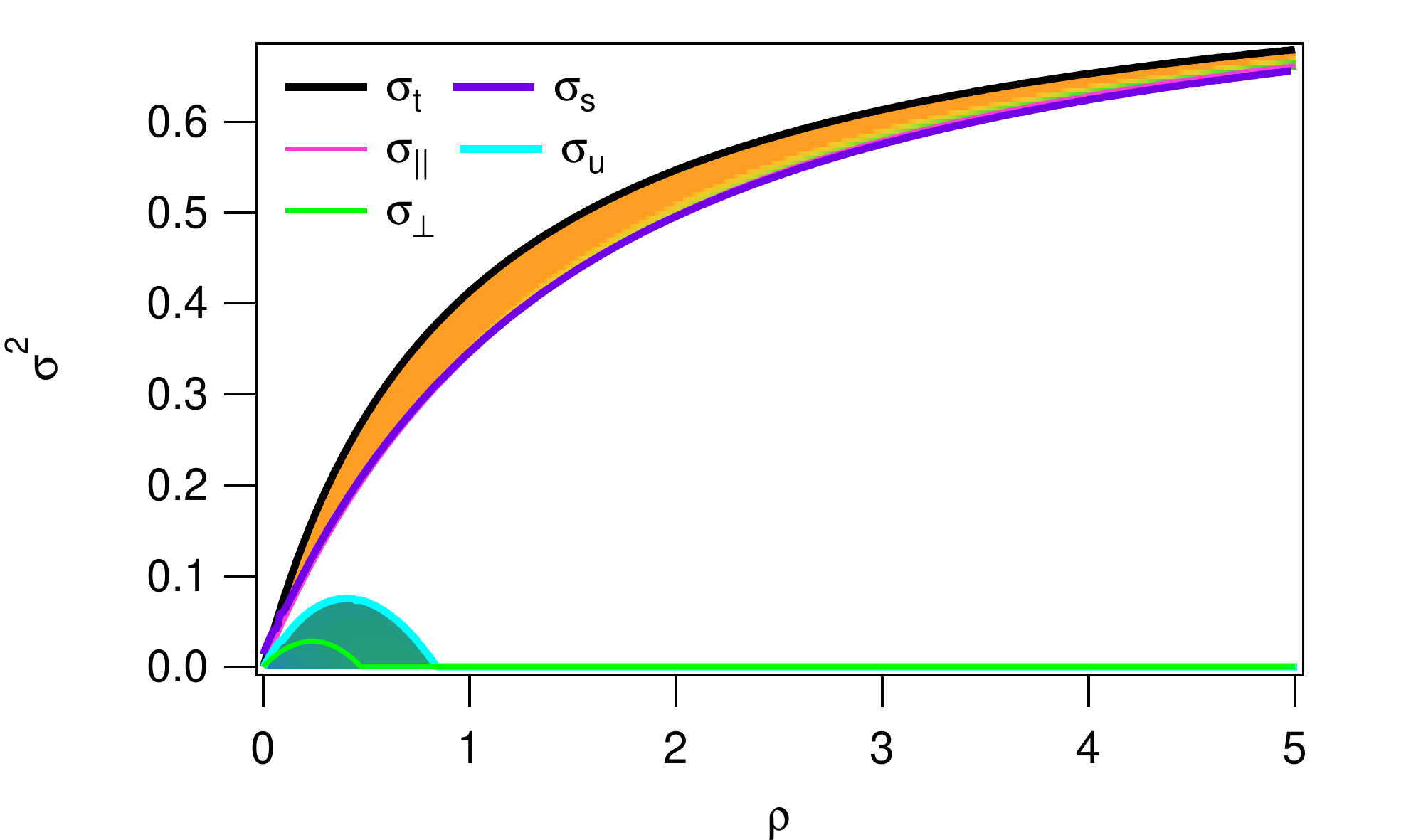}
\hfill
\includegraphics[width=0.5\columnwidth]{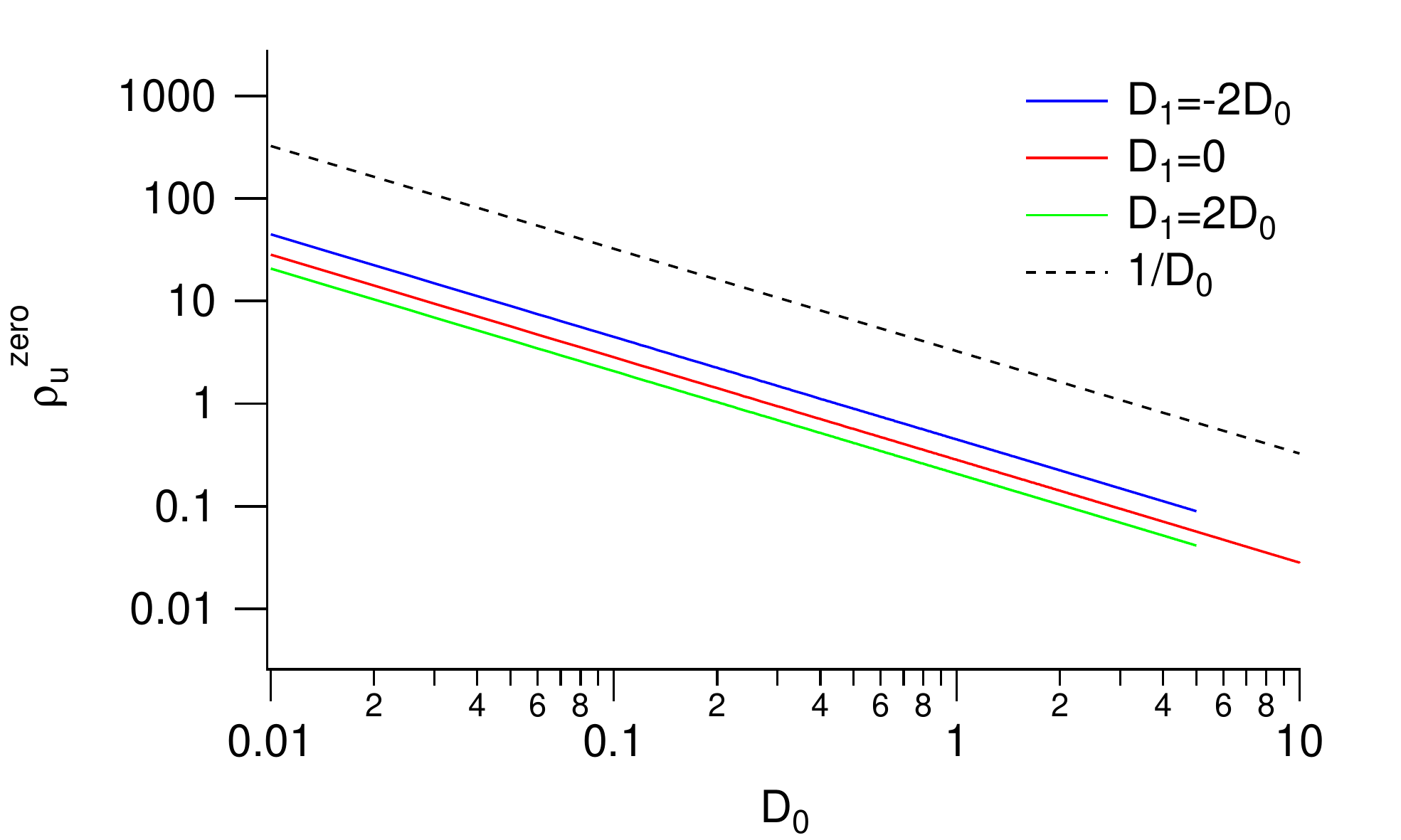}
\caption{
Phase diagram of the linear stability analysis of the homogeneous solutions to the hydrodynamic equations
for the case of polar particles aligning ferromagnetically in the presence of positional diffusion: $D_1=-0.5$ (top left), $D_1=0$ (top right), $D_1=0.5$ (bottom left); $D_0=0.25$ in all three cases, same colormap as in Fig.~\ref{diag-stab-polar}. 
Bottom right: intersection $\rho_u^{\rm zero}$ of the line $\sigma_u$ with the horizontal axis, as a function of $D_0$ for different cases (see legends), showing that 
$\rho_u^{\rm zero} \propto 1/D_0$.}
\label{diag-stab-diff}
\end{figure}

A positive $D_0$ shrinks a bit the first instability region while making its unstable modes more clearly longitudinal (see Fig.~\ref{diag-stab-diff}, top right, where $D_1=0$).
For any fixed positive $D_0$, the extent of the second region varies greatly with $D_1$: it can be made smaller for positive
$D_1$ values (Fig.~\ref{diag-stab-diff}, bottom left), and larger for negative $D_1$ values (Fig.~\ref{diag-stab-diff}, top left). We find that the size of the second region, as measured
by the location of the point $\rho_u^{\rm zero}$ where $\sigma_u$ intersects the horizontal axis,
actually scales exactly like $1/D_0$ (Fig.~\ref{diag-stab-diff}, bottom right).


\subsection{Inhomogeneous solutions and dynamics}

The linear stability analysis recalled above revealed, for metric models, the existence of a region
bordering the linear onset of order inside which the homogeneous ordered solution is unstable. Our hydrodynamic equations
then exhibit nonlinear, inhomogeneous, phase-separated solutions, which are faithful to those observed in microscopic simulations.
In our opinion, this makes a strong case in favor of our approach, which appears to yield an overall very good qualitative agreement
with relevant active particle models.

The inhomogeneous solutions take different form depending on the case considered: in the polar case, the hydrodynamic equations typically 
show trains of propagating ``bands", as explained recently in \cite{BDG,BARTOLO} and observed in microscopic models. The stability and 
selection mechanisms at work in these smectic arrangements is still under study. In the active nematics and self-propelled rods cases,
a dominating solution was observed and found analytically, which takes the form of a stationary ordered domain elongated along nematic order, and occupying a well-defined fraction of space \cite{ACTIVE-NEMA,RODS-KINETIC}. For active nematics, it was recently shown that this band solution is actually itself always linearly unstable, and leads to a large-scale spatiotemporally chaotic regime \cite{NEMA-CHAOS}.

Even though the phase-separated solutions take different forms depending on the
case considered, we find that they exist, and are observed, in a parameter region {\it wider} than the linear instability 
domain of the homogeneous state. This important observation immediately implies that there exist two coexistence regions (in the sense that two different solutions may exist for the same values of the parameters) at the 
level of the deterministic hydrodynamic equations. In particular the nonlinear phase-separated solutions coexist with the disordered phase, 
which explains the discontinuous character of the transition, and points to the ``irrelevance" of linear stability thresholds in the presence of
fluctuations. Indeed, they are much like spinodal lines in usual phase separation phenomena, i.e. they are {\it not} the transition 
points observed in systems with fluctuations.

\subsection{On the Boltzmann-Ginzburg-Landau approximation}

\subsubsection{Boltzmann equation}

The Boltzmann approach is based on the binary collision approximation
and the molecular chaos hypothesis. 
Under these conditions, the Boltzmann equation (\ref{FullBoltzmann}) faithfully
describes our microscopic dynamics on timescales $\tau_B$ much larger
than the microscopic dissipative scale $\Delta t$ and the collision
timescale $\tau_{\rm coll}$, and on lengthscales $\ell_B$ much larger
than the different microscopic scales, namely the elementary displacement $d_0=v_s \Delta t$
(where $v_s$ is the characteristic speed of particles), the interaction range $r_0$,
and the typical distance between particles $\ell \sim 1/\sqrt{\rho_0}$.
Metric-free models are briefly
discussed in Section \ref{metric-free}, so that here we will only discuss
metric models.

In metric models, binary collisions dominate the dynamics when
(i) the interaction range $r_0$ is much smaller then
the mean particle interdistance $\ell\sim1/\sqrt{\rho_0}$ (in two spatial
dimensions) and (ii) the intercollision time $\tau_{\rm free}$ is much
larger than the collision timescale $\tau_{\rm coll}$. 

In self-propelled systems, dominated by drift, one has 
\begin{equation}
\tau^{(b)}_{\rm free} \approx \frac{1}{v_0 \sqrt{\rho_0}}
\end{equation}
while in purely diffusive systems we get
\begin{equation}
\tau^{(d)}_{\rm free} \approx \frac{1}{D \rho_0}
\end{equation}
where $v_0$ is the drift speed and $D$ a positional diffusion constant (either
isotropic or anisotropic).
The collision time, on the other hand, is given by the time needed by
two interacting particles to separate; since interacting particles are
aligned up to some noise, it is always diffusion dominated 
\begin{equation}
\tau_{\rm coll} \approx \frac{r_0^2}{D} .
\end{equation}

In general the condition $\tau_{\rm coll} \ll \tau_{\rm free}$ (as well
as $r_0 \ll 1/\sqrt{\rho_0}$) can be fulfilled at low densities in both
ballistic and diffusive systems, thus justifying the binary collision
approximation in drift-diffusion dry active systems. 

It should be noted, however, that in a
 ballistic system like the original VM, positional diffusion is
strictly zero, and interacting particles may be only separated by
angular noise, with an effective diffusion constant
\begin{equation}
D_{\rm eff} \sim \frac{d_0^2 \eta^2}{\Delta t} .
\label{Deff}
\end{equation}
Since in metric systems the critical noise amplitude $\eta_c$ corresponding to the onset
of order decreases with density, this casts some doubts on the
applicability range of the binary collision approximation for the original VM in the absence of positional
diffusion. In particular, using a mean-field estimate \cite{CHATE} of the critical
noise $\eta_c \sim d_0 \sqrt{\rho_0}$, Eq.~(\ref{Deff}) becomes 
\begin{equation}
\tau_{\rm coll}^{(VM)} \approx \Delta t \frac{r_0^2}{d_0^4 \rho_0} 
\end{equation}
so that, with $v_0=d_0/\Delta t$, the condition $\tau_{\rm coll} \ll \tau_{\rm free}$ would imply
\begin{equation}
\frac{r_0^2 }{d_0^3 \sqrt{\rho_0}} \ll 1
\label{VM1}
\end{equation}
to be satisfied simultaneously with condition (i)
\begin{equation}
r_0 \ll \frac{1}{\sqrt{\rho_0}} .
\label{VM2}
\end{equation}
Eqs.~(\ref{VM1})-(\ref{VM2}) can only be satisfied simultaneously if
$(r_0/d_0)^3 \ll 1$, so that one expects that for the classic VM, as well
as for other ballistic models, higher-order collisions will
become relevant even at low densities for moderate to low speeds. 

The molecular chaos hypothesis, which allows one to decompose many-particle distributions into products of single-particle ones, on the other hand, is a rather
delicate issue.
It was investigated in Refs.~\cite{Degond}, using
a binary collision VM, through the chaos propagation property: if the latter holds, initially
decorrelated particles will stay decorrelated at future times, and thus
the molecular chaos assumption will be justified. The analysis of
\cite{Degond} is however inconclusive. While at the kinetic timescale
chaos propagation holds in the thermodynamic limit, longer time
behavior remains out of reach of exact methods (while preliminary numerical
simulations seem to indicate a violation of the molecular chaos hypothesis).  
See also \cite{Frey2013b} for related discussions using a numerical analysis.

\subsubsection{Convergence issues in the hydrodynamic expansion}

The second approximation involved in the BGL approach is the small
$\epsilon$ expansion used to close the hierarchy (\ref{eq-fk}) in
order to derive hydrodynamic equations. While this expansion is `exact'
at the onset of order, one can wonder whether the obtained hydrodynamic
equations remain valid away from the transition line, in
particular for what regards linear stability results.

For polar particles, in the explicit examples where the stability
analysis of the ordered phase has been performed in detail, we have
found an additional instability deeper in the ordered
phase, further away from the linear instability threshold of the disordered
phase (see sect.~\ref{sect-general-scenario}) \cite{TOPOKINETIC,RODS-KINETIC}. 
Such an additional instability was however not found for active nematics
\cite{ACTIVE-NEMA}. 

This instability is considered as spurious in the sense that it is not
observed in the corresponding microscopic models. It typically
occurs with very large growth rates, raising doubts as to its physical meaning.
It is likely to be an artifact of the truncation procedure. Though we have
not performed a detailed analysis of all three possible classes,
results obtained for polar particles with ferromagnetic interactions
indicate that this spurious instability is not present at the
Boltzmann level \cite{Unpublished}.

The convergence of the $\epsilon$-expansion of the Boltzmann equation
at a finite distance from the instability threshold is a complicated issue. A first
guess is that going to higher orders in the $\epsilon$-expansion could
improve the description further away from threshold, e.g. by removing, or at least shrinking,
the region of the additional, ``spurious" instability.

We have derived the hydrodynamic equations up to order 7 in $\epsilon$ for the canonical case of polar
particles aligning ferromagnetically (Vicsek model) \cite{ANTON_PHD}. 
It is relatively easy to see that only odd orders in $\epsilon$ need to be considered, and that even orders
yield intrinsically unbalanced equations. For example, at order 4, one keeps the equation for $f_2$, while $f_3$ is slaved and 
expressed as a function of $f_1$ and $f_2$, yielding equations of the form:
\begin{eqnarray}
\partial_t f_1 &=& a_{1} f_1 + a_{21} f_2 f_1^*  \\
\partial_t f_2 &=& b_{2} f_2 + b_{11} f_1^2 + b_{211} f_2 |f_1|^2 
\end{eqnarray}
(for clarity, terms involving space derivatives are not shown).
The last term of the second equation is the only new, fourth-order one. It can be shown
to render the homogeneous ordered solution linearly unstable everywhere.

At order 5, we obtain 3 equations of the form:
\begin{eqnarray}
\partial_t f_1 &=& a_{1} f_1 + a_{21} f_2 f_1^* + a_{23} f_2^* f_3  \\
\partial_t f_2 &=& b_{2}f_2 + b_{11} f_1^2 + b_{211} f_2 |f_1|^2  \\
\partial_t f_3 &=& c_{3} f_3 + c_{12} f_1 f_2 + c_{113} |f_1|^2 f_3 + c_{122} f_1^* f_2^2 
\end{eqnarray}
in which several order 5 terms appear. The homogeneous ordered solution now has to be determined 
numerically. We find that it exists only for a limited range of noise values, whereas no such limitation occurs 
at order 3. Thus order 5 is ``worse" than order 3.

At order 7, we now have 5 equations of the form:
\begin{eqnarray}
\partial_t f_1 &=& a_1 f_1 + a_{21} f_2 f_1^* + a_{23} f_2^* f_3 + a_{34} f_3^* f_4  \\
\partial_t f_2 &=& b_2 f_2 + b_{11} f_1^2 + b_{211} f_2 |f_1|^2 + b_{24} f_2^* f_4  \\
\partial_t f_3 &=& c_3 f_3 + c_{12} f_1 f_2 + c_{14} f_1^* f_4 + c_{25} f_2^* f_5  \\
\partial_t f_4 &=& d_4 f_4 + d_{22} f_2^2 + d_{13} f_1 f_3 + d_{15} f_1^* f_5   \\
\partial_t f_5 &=& e_5 f_5 + e_{14} f_1 f_4 + e_{23} f_2 f_3 + e_{115} |f_1|^2 f_5
+ e_{133} f_1^* f_3^2 + e_{124} f_1^* f_2 f_4 \, .
\end{eqnarray}
Now the homogeneous solution exists over the whole noise range again. 
The maximal value of polarization ($|f_1|/\rho$) reached at zero noise is higher than at order 3, 
which is an improvement. However, the linear stability study of the homogeneous ordered solution
shows that it is linearly unstable everywhere: even at order 7, the situation is less satisfactory than
at order 3.

Even though these results would need to be completed by a similar study performed on the altogether 
slightly better-behaved case of active nematics, they indicate already that the BGL approach only makes sense
at the usual, third-order, Ginzburg-Landau level. Exploring higher orders seems to only reveal that an angular Fourier 
expansion is not suited far from the onset of order, calling for more sophisticated expansion schemes, possibly along the lines
of the ``reductive perturbation theory" approach introduced by Chen et al. \cite{GOLDENFELD}.

\subsubsection{Extended Particles}

Our framework here has been restricted to interactions that respect the particle exchange symmetry.
However, it has been pointed out that `more realistic' extended particles may show a more
complicated collision dynamics \cite{Frey2013}. In particular,
collisions may locally
violate the particle exchange symmetry which, in a Boltzmann approach, can only be restored in a statistical sense.
In particular, Ref. \cite{Frey2013} computes the effective Boltzmann level collision dynamics for two different extended particles models:
self-propelled hard rods and a bead-spring model. In both cases, it is claimed that binary collisions do not lead to a positive linear coefficient $\mu_1$ for the polar order parameter. Contrary to the conclusions of Ref. \cite{Frey2013}, we do not find these results significative
to the point of constituting "a critical assessment" of the BGL approach, as there is no indication that the two microscopic models studied
give rise to polar order on the mesoscopic scale. On the other hand, our discussion in Sec. \ref{Sec:alignment} clearly shows that at sufficiently low noise, any interaction rule being indeed an alignment rule results, at the Boltzmann level, in the linear instability of the disordered solution towards the order parameter associated to the interaction symmetry.

\subsection{Comparison with other approaches}

Other statistical physics approaches have also been proposed to derive hydrodynamic equations for active particles, mainly the mean-field Fokker-Planck equation approach \cite{MCM-rod1,MCM-rod2,PAWEL,FARRELL} and the Enskog-like approach \cite{IHLE}.

\subsubsection{Mean-field Fokker-Planck equation}

In the mean-field Fokker-Planck equation case, one starts from interacting active particles described by a set of Langevin equations, from which a $N$-body Fokker-Planck equation is derived. This $N$-body equation is then treated in a mean-field way, meaning that the force experienced by a given particle is replaced by a local statistical average of the force. The $N$-body linear Fokker-Planck equation is then turned into a one-body nonlinear Fokker-Planck equation, under an assumption of factorization of the $N$-body distribution as a product of one-body distributions \cite{MCM-rod1,MCM-rod2,PAWEL,FARRELL}. If the active particles are described by underdamped Langevin equations, some approximation scheme is further needed to eliminate the velocity degrees of freedom, then reducing the mean-field Fokker-Planck to a Smoluchowski equation \cite{MCM-rod1}.

Although the starting points are significantly different, the resulting Smoluchowski (or Fokker-Planck in the overdamped case) equation bears some formal similarities with the Boltzmann equation. It also describes the evolution of the one-body distribution $f({\bf r},\theta,t)$, and takes the form of a bilinear integral equation. As a result, the angular Fourier expansion takes a form close to that obtained from the Boltzmann equation, and the same approximation scheme can be used close to the instability threshold of the disordered solution (though most papers following this approach use truncations at some lower order than we do in the BGL approach).

For point-like self-propelled polar particles with overdamped Langevin dynamics and ferromagnetic interactions, the resulting hydrodynamic equations are very similar to the ones obtained from the Boltzmann equation \cite{FARRELL}. 

For rods, the hydrodynamic equations derived from a Smoluchowski equation in \cite{MCM-rod1} are different from the ones we obtain here for polar particles with nematic interactions, for two main reasons. First, in \cite{MCM-rod1}, only $f_1$ and $f_2$ have been retained in the truncation, while we have kept modes up to $f_4$, and used $f_3$ and $f_4$, once enslaved to $f_1$ and $f_2$, to close  the hierarchy of equations. We thus have more terms (and in particular more non-linear terms, including the cubic term saturating the instability of the state $f_2=0$) in our equations.
Second, the spatial extension of the rods has been explicitly taken into account in \cite{MCM-rod1}, leading to the emergence of new terms coupling the polar order parameter $f_1$ to its gradient, in the equation for $f_1$, while we have considered point-like particles in our Boltzmann equation. Additional terms are thus obtained in \cite{MCM-rod1} with respect to our results.
More generally, the Fokker-Planck (or Smoluchowski) equations constitute a well-suited framework to include position- and/or orientation-dependent forces, as done in \cite{MCM-rod1} for extended rods, or in \cite{PAWEL2} for forces depending on the relative orientation of particles. Although we have not done it explicitly up to now, such extensions should also be possible in the context of the Boltzmann equation, and are expected to yield the same type of additional terms into the resulting hydrodynamic equations.

On the other hand, one of the drawbacks of the Fokker-Planck approach is its sensitivity to some details of the dynamics. Indeed, in the case of point-like self-propelled particles with overdamped Langevin dynamics considered in \cite{FARRELL}, it can be seen rather easily that normalizing or not the local force (exerted by neighboring particles) by the number of neighbors results in significantly different macroscopic dynamics at the level of the hydrodynamic equations, since the linear term in the equation for the polar order parameter becomes independent of the density if the force is normalized --recall that, in this case, the basic instability of the homogeneous ordered
state is suppressed (as in metric-free models).

\subsubsection{Enskog equation approach}
\label{enskog}

An alternative method has been proposed, based on an Enskog-like equation, in order to deal with multiple collisions in the semi-dilute regime \cite{IHLE}. Its main advantage, with respect to the Boltzmann approach, is that it is not limited to binary interactions; all multiple interactions are taken into account. However, it also relies, like the other methods, on a factorization assumption of the $N$-body distribution fonction, or in other words, on a 'molecular chaos' assumption.
From the Enskog equation, an angular Fourier expansion can also be performed.
In principle, an $\epsilon$-expansion close to the instability threshold could be performed, in the same way as for the Boltzmann equation. The derivation made in \cite{IHLE} is presented in a slightly different way. In particular the truncation and closure performed leads to
hydrodynamic equations with many more (higher-order) terms than those retained in the BGL approach. At the level of the linear stability 
of the homogeneous solutions, these equations yield more quantitatively accurate thresholds than BGL, as compared to numerical simulations of the VM. This is satisfactory in principle,
but in view of the generic presence of the long-wavelength instability of the ordered state leading to bands, and of the existence and stability domains of these nonlinear solutions extending beyond the linear thresholds, the transition lines observed in the presence of fluctuations 
(and firstly in the microscopic VM) cannot be the linear thresholds found in the (deterministic) hydrodynamic equations.
Moreover, at the nonlinear level, the many terms retained in \cite{IHLE} seem to lead to unbounded solutions: the obtained hydrodynamic equations are ill-behaved.

\section{Conclusion}

In this paper, we have presented a generic Boltzmann-Ginzburg-Landau framework allowing us to derive, in a controlled way, continuous equations describing the large-scale behavior of assemblies of point-like active particles interacting through polar or nematic alignment rules.
Also varying the symmetry properties (polar or nematic) of the
interactions between particles, we have identified three main classes of dry active systems, and derived for each class the corresponding continuous equations for the relevant order parameters.
One of the main advantages of our approach is that the truncation and
closure of the infinite hierarchy is well-controlled, using standard
scaling arguments close to the linear instability threshold of the
disordered state. As often the case for Ginzburg-Landau-type
equations, the resulting minimal equations actually also describe, in
a qualitative manner, the behavior of the system further away from the
linear instability threshold, when nonlinear structures appear and
possibly have a complicated spatiotemporal dynamics. We thus believe
that on the one hand, our truncation method to derive
continuous hydrodynamic equations is quite general, and can be applied to other starting points like Fokker-Planck or Smoluchowski equations \cite{FARRELL,PAWEL2}, and on the other hand that the resulting equations are, for each class, the minimal well-behaved equations accounting for the relevant phenomenology.

\subsection*{Acknowledgements}
We thank the Max Planck Institute for the Physics of Complex Systems, Dresden,
for providing the framework of the Advanced Study Group
``Statistical Physics of Collective Motion''
within which part of this work was conducted.
We also thank the Kavli Institute for Theoretical Physics
in Santa Barbara where this work was completed, within the framework of the program
"Active Matter: Cytoskeleton, Cells, Tissues and Flocks".
We are also grateful to A. Baskaran and M.C. Marchetti for useful discussions.
FG acknowledges support from grants
EPSRC First Grant EP/K018450/1 and MC Career Integration
Grant PCIG13-GA-2013-618399.






\end{document}